\documentclass[11pt, a4paper]{article}
\usepackage[latin1]{inputenc}
\usepackage[english]{babel}
\usepackage{amssymb}
\usepackage{amsmath}
\usepackage{amsthm}
\usepackage[]{graphicx}
\usepackage[]{subfigure}
\usepackage{tensor}
\usepackage{color}
\usepackage{cancel}
\usepackage{setspace}
\usepackage{fancyhdr}
\usepackage[bookmarks,linktocpage, colorlinks=true, plainpages = false, citecolor = blue,  linkcolor=blue, urlcolor = blue, filecolor = blue]{hyperref} 
\newcommand{\smallWidthLeft}{222.5pt}
\newcommand{\smallWidthRight}{212.5pt}

\newcommand{\hugeWidth}{435pt}

\title{\bf Spherical perturbations of hairy black holes in designer gravity theories}

\author{Lorenzo Battarra\footnote{lorenzo.battarra@apc.univ-paris7.fr}\\
{\small {}}\\
{\small {\it  APC (CNRS, Universit\'e
Paris-Diderot) , 10 rue Alice Domon et L\'eonie Duquet,}}\\
{\small {\it  75205 Paris Cedex 13, France}}\\
}

\begin{document}
%\date{\today}
\maketitle

\begin{abstract}
\noindent
We study the spectrum of scalar $l=0$ quasi--normal frequencies of Anti--de--Sitter hairy black holes in four and five dimensional designer gravity theories of the Einstein--scalar type, arising as consistent truncations of $ \mathcal{N} = 8$ gauged supergravity. In the dual field theory, such hairy black holes represent thermal states in which the operator corresponding to the bulk scalar field is condensed, due to the multi--trace deformation associated with non--standard boundary conditions. We show that, in a particular class of models, the effective potential describing the vacua of the deformed dual theory can be identified, at large values of the condensate, with the deformation plus the conformal coupling of the condensate to the curvature of the boundary geometry. In this limit, we show that the least damped quasi--normal frequency of the corresponding hairy black holes can be accurately predicted by the curvature of the effective potential describing the field theory at finite entropy.
\vspace{1.5cm}
\end{abstract}

\section{Introduction}
The last stage in the approach to a black hole geometry consists of a superposition of damped oscillations, whose complex frequencies are called quasi--normal frequencies. Following the seminal work \cite{Horowitz2000}, increasing efforts have been devoted to the computation of quasi--normal frequencies of \textit{``black''} geometries with AdS boundary conditions, motivated by the AdS/CFT correspondence. Indeed, when a dual field theory description is available, we know that these frequencies govern the approach to thermal equilibrium, appearing as poles of the finite temperature retarded correlators of the dual operator \cite{Birmingham2002, Son2002}.

In this paper, we study the quasi--normal frequencies of four and five dimensional black holes dressed with a single neutral scalar field, motivated by consistent truncations of $ \mathcal{N} = 8$ gauged supergravity. The scalar field mass squared  belongs to a particular range in which both asymptotic modes at the AdS boundary are normalizable, and one is free to impose general boundary conditions that mix their coefficients, denoted $( \alpha, \beta)$ in the following. The resulting Einstein--scalar theories are called \textit{designer gravity theories}, because their properties  depend on the free choice of a function that determines the boundary conditions. In the dual field theory, such function appears as a deformation of the action: this freedom made it possible to engineer holographic setups opening new perspectives in traditionally hard research topics, like the formation and possible resolution of cosmological singularities \cite{Hertog2004d, Craps2007}, and strongly coupled field theories at quantum critical points \cite{Faulkner2010b}.

Effective potentials built from supergravity data are an important tool in the analysis of these setups \cite{Hertog2005b}. They are computed from the coefficients $( \alpha, \beta)$ of the asymptotic modes of the scalar field that characterize families of hairy black hole solutions, and include the deformation which corresponds to the boundary conditions. Starting from the family of hairy black holes of a given radius $r_h$, one can build a function $ \mathcal{V}_{r_h}( \alpha)$\footnote{In the degenerate case $m^2 = m_{BF}^2$, $ \mathcal{V}_{r_h}$ is more conveniently expressed as a function of the other coefficient, denoted by $ \beta$ in the following}, whose extrema $ \alpha ^{*}$ correspond to the hairy black hole solutions of that radius that fulfill the boundary conditions. Moreover, $ \mathcal{V}_{r_h}( \alpha ^{*})$ gives the total mass of such solutions. In the dual field theory, $ \alpha$ represents the expectation value of the operator that condensates in the hairy black hole states, while the mass and the horizon area are naturally identified with the total energy and the entropy at equilibrium. Therefore, $ \mathcal{V}_{r_h}$ is interpreted as the effective field theory potential describing the thermodynamics at fixed entropy: the equilibrium states are extrema of the total energy. An analogous construction starting from the family of hairy black holes of given temperature $T$ leads to a function $ \mathcal{V}_{T}$ that is naturally interpreted as the effective field theory potential in the canonical ensemble at temperature T. Indeed, the extrema $ \alpha ^{*}$ correspond to the black hole solutions of temperature $T$ that fulfill the chosen boundary conditions, and $ \mathcal{V}_{T}( \alpha ^{*})$ gives the free energy of such states.

At the level of computing the energy/free energy of the equilibrium states, these effective potentials encode all the information about the strong interactions that characterize the dual field theory: they appear to provide a genuinely non--perturbative picture of arising from supergravity data. Previous studies qualitatively showed that they also give a correct intuition for the stability/instability of the equilibrium states under perturbations of the condensate, depending on the boundary conditions \cite{Hertog2005c}. Our work aimed at gene\-ralizing this observation, and verifying whether precise information about the quasi--normal frequencies of the hairy solutions can be extracted from the form of the effective potentials themselves.

In Section 2, we introduce the general bulk setup, consisting of a single neutral scalar field with negative mass squared coupled to Einstein gravity, and review the existence and properties of AdS static hairy solutions, as well as of the effective potentials. We also recall the definition of quasi--normal frequencies, briefly  describing the numerical procedure we have adopted. In Section 3, we present numerical results for five dimensional hairy black holes, choosing regularized boundary conditions of the type considered in \cite{Battarra2010}. In Section 4, we generalize our results to a four dimensional model that exhibits similar properties and, based on a further four dimensional model, we discuss the limitations of our approach to quasi--normal frequencies based on the effective potentials.
\section{Einstein--scalar setup}
\subsection*{Spherically symmetric, static solutions}
We start from the general Einstein--scalar action:
\begin{equation}\label{eq:action}
S_d  =  \frac{1}{8 \pi G_d} \int d^d x\, \sqrt{-g}\, \Big( \frac{R}{2} - \frac{1}{2} ( \partial \varphi)^2 - V( \varphi) \Big) \;.
\end{equation}
The potential $V$ is taken to have a negative local maximum in $ \varphi = 0$, corresponding to a unit radius AdS solution:
$$
V( \varphi) = - \frac{(d-1)(d-2)}{2} + m^2 \varphi^2 + \mathcal{O}( \varphi^3) \;.
$$
Denoting $m_{BF}^2$ the Breitnerlohner--Freedman bound, $
m_{BF}^2 = - (d-1)^2/4
$, we will consider models with $ m_{BF}^2 \leq m^2 < m_{BF}^2 + 1$, for which both the asymptotic modes of the scalar field are normalizable. The degenerate case $ m^2 = m_{BF}^2$ must be considered separately. In global coordinates the AdS metric reads:
$$
ds^2 = - (1+r^2) dt^2 + \frac{dr^2}{1 + r^2} + r^2 d \Omega_{d-2} \;,
$$
and the scalar field decays near the AdS boundary $r = \infty$ as:
\begin{eqnarray} \label{eq:asyScalar}
\begin{array}{ll}
m_{BF}^2 < m^2 < m_{BF}^2 + 1: \;& \varphi  =  \alpha\, r^{ -\lambda _{-}}+ \beta\, r^{ -\lambda _{+}} \;,\vspace{0.3cm}\\
m^2 = m_{BF}^2: & \varphi = \alpha\, r^{- \lambda} \log{ r} + \beta\, r^{- \lambda} \;,
\end{array}
\end{eqnarray}
$$
\lambda _{ \pm} = \frac{d-1}{2} \Big(1 \pm \sqrt{1 - m^2/m_{BF}^2} \Big) \;,
$$
In this range of masses, basic physical requirements such as the finiteness of the total energy do not require the slower decaying mode to vanish, $ \alpha = 0$. Therefore, we are free to consider more general boundary conditions, that are conveniently parametrized as:
\begin{eqnarray} \label{eq:bcs}
\begin{array}{ll}
m_{BF}^2 < m^2 < m_{BF}^2 + 1: \;& \beta = W_{, \alpha} \vspace{0.3cm}\\
m^2 = m_{BF}^2\; & \alpha = - W_{, \beta}
\end{array}
\end{eqnarray}
where $W( \alpha)$ (or $ W( \beta)$ in the degenerate case) is an arbitrary function. A particular solution to the field equations can be retained only if the corresponding values of $ \alpha$ and $ \beta$ satisfy \eqref{eq:bcs}. For a generic choice of $W$, some of the asymptotic AdS symmetries are broken. When the gravitational theory admits a field theory dual, the choice of boundary conditions \eqref{eq:bcs} is mapped into a deformation of the action of the boundary theory \cite{Witten2001, Vecchi2010b}.

Adopting a convenient gauge, the most general spherically symmetric solution of the field equations can be written as:
\begin{eqnarray}\label{eq:ansatz1}
ds^2 & = & - f(r,t) e^{-2 \delta(r,t)} dt^2 + \frac{dr^2}{f(r,t)} + r^2 d \Omega_{d-2} \;,\\ \label{eq:ansatz2}
\varphi & = & \varphi(r,t) \;.
\end{eqnarray}
The field equations stemming from \eqref{eq:action} read:
\begin{eqnarray*}
\partial _{t} \left( e^{ \delta} f ^{-1} \dot{ \varphi} \right) + \frac{1}{r ^{d-2}} \partial _{r} \left( r ^{d-2} e^{ - \delta} f \varphi' \right)  -   e^{ - \delta} V _{,\varphi} & = & 0 \;,\\
\dot{ f}  + \frac{2 r}{d-2} f \dot{ \varphi} \varphi' & = & 0 \;,\\
\delta'   + \frac{r}{d-2} \left( f ^{-2} e^{2 \delta} \dot{ \varphi}^2 + \varphi ^{\prime 2} \right) & = & 0 \;,\\
(d-3)(1-f) - r f' + r f \delta' -  \frac{2}{d-2} r^2 V  & = & 0 \;.
\end{eqnarray*}
In the static case, the field equations take the form:
\begin{eqnarray*}
f \varphi ^{\prime \prime} + \left( \frac{f}{r} + \frac{d-3}{r} - \frac{2 r V}{d-2} \right) \varphi' & = & V_{, \varphi} \;,\\
(d-3)(1-f) - r f' - \frac{r^2}{d-2} f \varphi ^{\prime 2} & = & \frac{2 r^2 V}{d-2} \;,\\
\delta' & = & - \frac{r \varphi ^{\prime 2}}{d-2} \;.
\end{eqnarray*}
A constant shift in $ \delta$ corresponds to a rescaling of the time coordinate, therefore $ \delta$ appears only through its derivative in the equations for static solutions. For this reason, we imply the gauge choice:
$$
\delta(r) \stackrel{r \rightarrow \infty}{ \longrightarrow} 0 \;,
$$
for which $ g_{tt}\, g_{rr} \stackrel{r \rightarrow \infty}{ \longrightarrow} -1$. Asymptotically, the field equations give:
\begin{eqnarray*}
\lambda _{+} \neq \lambda _{-}: \quad f & = & r^2 + 1 + \frac{ \lambda _{-} \alpha^2}{(d-2) r^{2 \lambda _{-} - 2}} - \frac{M_0}{r^{d-3}} + \ldots\;, \\
\lambda _{+} = \lambda _{-}: \quad f & = & r^2 + 1 + \frac{ \lambda \alpha^2 \log^2 r}{(d-2) r^{d-3}} + \frac{ \alpha( 2 \lambda \beta - \alpha) \log{r}}{(d-2)r^{d-3}} - \frac{M_0}{r^{d-3}} + \ldots\;,
\end{eqnarray*}
where $M_0$ is an integration constant and $ \alpha$, $ \beta$ are the constants appearing in \eqref{eq:asyScalar}. The most general solution of the static field equations with $ \varphi =0$ is of the Schwarzschild--AdS (SAdS) type:
\begin{eqnarray}
\label{eq:SAdS}f & = & 1 + r^2 - \frac{M_0}{r^{d-3}}, \quad M_0 = r_h^{d-3} + r_h^{d-1} \;, \\
\nonumber \delta & = & 0 \;.
\end{eqnarray}
However, for a variety of potentials $ V( \varphi)$ and spacetime dimensionalities, both analytical and numerical black hole solutions have been found for which the scalar field has a non--vanishing profile, and $ \varphi \rightarrow 0$ at $ r \rightarrow \infty$.  For a given potential, these solution generally form a two--parameter family spanned by the values $ \varphi_h$ of the scalar field at the horizon and of the horizon radius:
$$
f( r_h) = 0, \quad \varphi(r_h) \equiv \varphi _{h} \;.
$$
Alternatively, each solution is uniquely characterized by the $( \alpha, \beta)$ coefficients of the scalar field, and the field equations provide a mapping $(r_h, \varphi_h) \leftrightarrow ( \alpha, \beta)$. Varying $ \varphi_h$ and keeping $ r_h$ fixed, one obtains a curve $ \beta_{r_h}( \alpha)$ (or $ \alpha_{r_h}( \beta)$) in the $ (\alpha, \beta)$ plane, that is generally not single--valued. Finally, hairy solitons complete the ensemble of static, spherically symmetric solutions to \eqref{eq:action} with $ \varphi \neq 0$. These are defined by the requirement of regularity at $ r=0$, and form a one--parameter family spanned e.g. by the value of the scalar field at the origin:
$$
\varphi _{h} \equiv \varphi(r=0)\;.
$$
Correspondingly, the integration of the field equations provides a single $ \beta_s( \alpha)$ curve (or $\alpha_s( \beta)$) describing the asymptotic parameters of the hairy solitons. In all the examples we have considered, one recovers $ \beta_s( \alpha)$ (resp. $ \alpha_s( \beta)$) as the $ r_h \rightarrow 0$ limit of $ \beta_{r_h}( \alpha)$ (resp. $ \alpha_{r_h}( \beta)$).

The choice of boundary conditions at infinity restricts the ensemble of admissible hairy solutions to a one parameter family. For example, the number of admissible hairy black holes of radius $ r_h$ is given by the number of intersections of the boundary conditions curve $ W_{, \alpha}( \alpha)$ with $ \beta_{r_h}( \alpha)$ or, in the degenerate case, of $- W_{, \beta}( \beta)$ with $\alpha_{r_h}( \beta)$.

Asymptotic time translational symmetry implies the conservation of the total mass \cite{Balasubramanian1999}, whose expression receives a finite contribution from the slow falloff of the scalar field:
\begin{eqnarray} \label{eq:mass1}
\lambda _{+} \neq \lambda _{-}: \quad M & = & \frac{V_{S^{d-2}}}{8 \pi G_d} \left( \frac{(d-2)}{2} M_0 + \lambda _{-} \alpha \beta + ( \lambda _{+} - \lambda _{-}) W ( \alpha) \right) \;,\\ \label{eq:mass2}
\lambda _{+} = \lambda _{-}: \quad M & = & \frac{V_{S^{d-2}}}{8 \pi G_d} \left( \frac{(d-2)}{2} M_0 + \frac{ \lambda}{2} \beta^2 + W( \beta) \right) \;.
\end{eqnarray}
The boundary conditions are insensitive to a constant shift in the value of $W$: the common choice for boundary conditions that allow non--hairy solutions ($ W'(0) = 0$) is to set $ W(0)=0$, in such a way that the mass of pure AdS vanishes. Formulas \eqref{eq:mass1}, \eqref{eq:mass2} can be straightforwardly generalized to the case in which the boundary conditions curve can be put into the form \eqref{eq:bcs} only piece--wisely: a suitable integration constant must be chosen on each piece, in such a way that $W$ is continuous along the curve itself.

As suggested in \cite{Hertog2005a}, for any $r_h$ one can build a constant entropy effective potential $ \mathcal{V}_{r_h}$:
\begin{eqnarray}
\label{eq:effPotNeq}
\lambda _{+} \neq \lambda _{-}: \quad  \mathcal{V}_{r_h}( \alpha) & = & \frac{M_{SAdS}(r_h)}{V_{S^{d-2}}} + \frac{ \lambda _{+} - \lambda _{-}}{8 \pi G_d}\left(W( \alpha) - \int_0 ^{ \alpha} \beta_{r_h} d \alpha' \right) \;, \\
\label{eq:effPotEq}
\lambda _{+} = \lambda _{-}: \quad \mathcal{V}_{r_h}( \beta) & = & \frac{M_{SAdS}(r_h)}{V_{S^{d-2}}}+ \frac{1}{8 \pi G_d} \left( W( \beta) + \int_0^ \beta \alpha_{r_h} d \beta' \right) \;,
\end{eqnarray}
where $M_{SAdS}$ is the mass of the SAdS solution of the same radius:
$$
M_{SAdS} = V_{S^{d-2}} \frac{(d-2)(r_h^{d-3}+r_h^{d-1})}{16 \pi G_{d}} \;.
$$
Each extremum $ \alpha ^{*}$ (resp. $ \beta ^{*}$) of $ \mathcal{V}_{r_h}$ corresponds to a hairy black hole of radius $ r_h$ satisfying the boundary conditions set by $ W$. Moreover, the total mass of such black hole is simply given by $ V_{S^{d-2}} V_{r_h} (\alpha ^{*})$ (resp. $ V_{S^{d-2}} V_{r_h} (\beta ^{*}))$. An effective potential $ \mathcal{V}_{s}$ can be built out of the soliton curve in the same way:
\begin{eqnarray}
\label{eq:effPotsNeq}
\lambda _{+} \neq \lambda _{-}: \quad  \mathcal{V}_{s}( \alpha) & = & \frac{ \lambda _{+} - \lambda _{-}}{8 \pi G_d}\left(W( \alpha) - \int_0 ^{ \alpha} \beta_{s}\, d \alpha' \right) \;, \\
\label{eq:effPotsEq}
\lambda _{+} = \lambda _{-}: \quad \mathcal{V}_{s}( \beta) & = & \frac{1}{8 \pi G_d} \left( W( \beta) + \int_0^ \beta \alpha_{s}\, d \beta' \right) \;.
\end{eqnarray}
$ \mathcal{V}_s$ plays a key role in giving the energy of the ground states: in \cite{Faulkner2010}, the authors have proven that the total energy in the theory \eqref{eq:action} \textit{with boundary conditions} \eqref{eq:bcs} is bounded from below if and only if $ \mathcal{V}_s$ has a global minimum. If this is the case, the minimum energy solution is precisely the hairy soliton that corresponds to the minimum of $ \mathcal{V}_s$.

\subsection*{Scalar, spherically symmetric perturbations of the ``hair''}
We now turn to the possibility to extract additional information from the shape of the effective potentials $ \mathcal{V}_s$ and $ \mathcal{V}_{r_h}$. We start from a very simple idea, namely that some perturbative properties of the hairy solutions may be encoded in the form of the effective potential near its extrema. For this reason, we focus on the scalar perturbations of the hairy black holes and solitons. In \cite{Hertog2005c}, the authors have computed the frequency of growing modes  of four and five dimensional unstable hairy black holes. In the remaining sections, we generalize and complete their results to models and boundary conditions that also admit perturbatively stable black hole solutions, focusing on the dependence of such frequencies on the parameters that specify the boundary conditions.

Due to the non--zero vev of the scalar field $ \varphi$, its perturbations give a backreaction on the metric already at the linear order. We limit our analysis to the case of spherically symmetric $ l = 0$ perturbations, that fit in the non--linear ansätz \eqref{eq:ansatz1}, \eqref{eq:ansatz2}: in this case, the linearized field equations can be reduced to a single master equation for the scalar field $ \varphi$. Therefore, we consider the linearized ansätz:
\begin{eqnarray}\label{eq:linearAnsatz1}
f & = & f_0(r) + \epsilon\, F(r) \,e^{- i \omega t} \;,\\ \label{eq:linearAnsatz2}
\varphi & = & \varphi_0(r) + \epsilon\, r^{1 - d/2}\, \Phi(r) \,e^{- i \omega t} \;,\\ \label{eq:linearAnsatz3}
\delta & = & \delta_0(r) + \epsilon\, \Delta(r)\, e^{-i \omega t} \;.
\end{eqnarray}
Except for static perturbations, that describe infinitesimal displacements along the one--parameter family of hairy black hole solutions, the linearized Einstein equations let one express $F$ and $ \Delta'$ in terms of $ \Phi$:
\begin{eqnarray*}
F & = & - \frac{2r f_0 \varphi_0 '}{d-2} \Phi \;,\\
\Delta' & = & - \frac{2r \varphi'_0}{d-2} \Phi' \;.
\end{eqnarray*}
The first of these equation can be used to extract the time--dependent variation of $M_0$ as a function of the variations of $ \alpha$ and $ \beta$, and check that the total mass \eqref{eq:mass1}, \eqref{eq:mass2} is stationary around static solutions. Introducing the tortoise coordinate $dr ^{*} \equiv f_0^{-1} e^{ \delta_0} dr$, the master equation takes the familiar Schrödinger form:
\begin{equation} \label{eq:perturbations}
\frac{d^2 \Phi}{d r ^{* 2}} - f_0 e^{-2 \delta_0} V_0 \Phi  =  - \omega^2 \Phi \;, \vspace{-0.3cm}
\end{equation}
\begin{eqnarray*}
V _0 & = & V _{, \varphi \varphi}  + \frac{4 r}{d-2} \varphi'_0 V _{, \varphi} + f_0 \left( \frac{(d-2)(d-4)}{4 r^2}  - \frac{3d - 10}{2(d-2)} \varphi _0^{\prime 2} - \frac{2 r^2}{(d-2)^2} \varphi_0 ^{\prime 4} \right) \\ && -
f_0' \left( \frac{2 r}{d-2} \varphi_0 ^{\prime 2} - \frac{d-2}{2 r} \right) \;.
\end{eqnarray*}
Near the horizon, located at $r ^{*} = - \infty$, $ f_0$ vanishes exponentially, and the field variable satisfies a free equation:
\begin{equation} \label{eq:nearHorizon}
\Phi = \Phi_{in}\, e^{- i \omega r ^{*}}(1 + \mathcal{O}(e^{2 \kappa r ^{*}})) + \Phi _{out}\, e^{ i \omega r ^{*}}(1 + \mathcal{O}(e^{2 \kappa r ^{*}})) \;,
\end{equation}
where $ \kappa \equiv f_0'(r_h) e^{- \delta_h}/2$ is the surface gravity of the hairy black hole. Quasinormal frequencies are defined as the values of $ \omega$ such that the master equation \eqref{eq:perturbations} admits a solution that is purely ingoing at the horizon ($ \Phi _{out} = 0$), and satisfies the boundary conditions at infinity, where the field perturbation decays as:
\begin{eqnarray*}
\lambda _{-} \neq \lambda _{+}: \quad \Phi & = &  A\, r^{-\lambda _{-} - 1 + d/2} + B \, r^{ - \lambda _{+} - 1 + d/2} \;,\\
\lambda _{-} = \lambda _{+}: \quad \Phi & = & A\,  r^{-\lambda - 1 + d/2} \log{r} + B \, r^{ - \lambda - 1 + d/2} \;.
\end{eqnarray*}
At the linearized level, the boundary conditions at the AdS boundary \eqref{eq:bcs} read:
\begin{eqnarray}\label{eq:linBcs1}
\lambda _{-} \neq \lambda _{+}: \quad B & = & W_{,  \alpha \alpha} (\alpha_0)\, A \;,\\ \label{eq:linBcs2}
\lambda _{-} = \lambda _{+}: \quad A & = & - W_{,  \beta \beta} (\beta_0)\, B \;.
\end{eqnarray}
Normal frequencies of hairy solitons are studied in the same way: ingoing boundary conditions at the horizon are replaced by the requirement of regularity at the origin. The corresponding Schrödinger problem is hermitian, and the eigenvalues $ \omega^2$ need to be real. Therefore, in the absence of numerical issues, one finds only pure real or imaginary frequencies. In all the examples we have studied, as already observed for SAdS black holes \cite{Horowitz2000} and many others, these normal frequencies coincide with the $r_h \rightarrow 0$ limit of the quasi--normal frequencies characterizing the hairy black holes which correspond to the solitons (and compatible with the same boundary conditions).

In the models we have considered, the background hairy black holes are known only numerically. Therefore, the generalization of the recurrence relation technique \cite{Horowitz2000} inspired by the polynomial form of the SAdS metric \eqref{eq:SAdS} involves a recurrence containing infinite terms at each step. One needs to truncate not only the number of iterations, but the number of terms involved at each iteration \cite{Zhidenko2006}, and the very interest of the original procedure is essentially lost. For this reason, our numerical technique consisted in a direct numerical resolution of the Schrödinger equation by shooting both from the horizon and the AdS boundary, imposing the relevant boundary conditions: ingoing propagation and \eqref{eq:linBcs1}, \eqref{eq:linBcs2}, respectively. Via a local optimization algorithm, one explores the complex plane of $ \omega$ until the corresponding solution satisfies the correct boundary condition at the opposite end of the resolution interval (\eqref{eq:linBcs1}, \eqref{eq:linBcs2} and ingoing propagation, respectively). Solving for $ \omega$ in these two independent ways provides a solid consistency check to the numerical computations.

As clearly explained in \cite{Kaminski2010a}, the shooting method is severely limited in determining quasi--normal frequencies with imaginary part much larger  than $ \kappa$ in absolute value. This is typically the case for high overtones, and is due to the fact that the corresponding modes vanish rapidly near the horizon. Going back to the usual radial variable, near the horizon one has:
$$
\Phi = \Phi _{in} (r - r _{h}) ^{- i \omega/ 2 \kappa} \left( 1 +  \mathcal{O}(r-r_h) \right) + \Phi _{out} (r - r _{h}) ^{ i \omega/2 \kappa} \left( 1 +  \mathcal{O}(r-r_h) \right)\;.
$$
The larger negative $ \textrm{Im}( \omega)$, the more difficult is to resolve the presence of the outgoing mode near the horizon. To tackle this problem, the authors of \cite{Kaminski2010a} propose a resolution algorithm based on the relaxation method. Our work being oriented to the study of the parameter dependence, we have decided to implement the shooting algorithm with a few improvements:
\begin{itemize}
\item When shooting from the horizon, the outgoing mode must be eliminated by imposing boundary conditions on the field variable $ \Phi$ and its derivative $ \Phi'$. This results in two distinct problems. First, depending on the value of $ \textrm{Im}( \omega)/ \kappa$, the outgoing mode can be subdominant with respect to the subdominant terms of the ingoing mode. Therefore, we compute analytically, by deriving the master equation, a sufficiently large number of coefficients of such subdominant terms. To be precise, one must know the value of $n$ such coefficients, where:
$$
n = \lceil \frac{| \textrm{Im}(\omega)|}{\kappa} \rceil \;.
$$
Still, the numerical precision must be sufficient to resolve all such subdominant terms in the vicinity of the horizon. This turns to be the decisive constraint in the computation of high overtones.
\item When shooting from the boundary of AdS, one can fit the solution near the horizon in order to estimate $ \Phi_{out}$ and $ \Phi_{in}$. This method has the advantage not to require the evaluation of the coefficients of the subdominant terms of the ingoing mode: one can fit, besides than $ \Phi_{out}$ and $ \Phi_{in}$, the coefficients of a the power expansion near the horizon:
\begin{eqnarray*}
\Phi & = & \Phi_{in} (r-r_h)^{- i \omega/ 2 \kappa}\left( 1 + c_{1}(r-r_h) + c_{2}(r-r_h)^2 + \ldots + c_n(r-r_h)^n \right)  \\
&& + \Phi_{out} (r-r_h)^{i \omega/ 2 \kappa} + \mathcal{O}(r-r_h)^{n + 1 - | \textrm{Im}( \omega)|/ 2 \kappa } \;.
\end{eqnarray*}
However, the limitation imposed by the numerical precision of the procedure is always present.
\end{itemize}
Depending on the model and the boundary conditions, it has been possible to find the first two or three overtones with $n=1$ or $n=2$.

The repeated procedure consisting of solving the field equations, find the hairy solutions compatible with certain boundary conditions and then computing the quasi--normal frequencies that characterize them, has been implemented as a Mathematica library containing simple optimization algorithms and methods to manipulate the input/output. The library is adapted for general spacetime dimensionalities and models of the type \eqref{eq:action}, and is available from the author upon request.

\section{Hairy solutions in \texorpdfstring{$ D = 5$}{D = 5}, \texorpdfstring{$ \mathcal{N} = 8$}{N = 8} gauged supergravity}
First, let us consider the following model:
\begin{equation} \label{eq:5Dpotential}
S = \frac{1}{8 \pi G_5} \int d^5x\, \sqrt{ - g} \left( \frac{R}{2} - \frac{1}{2}( \partial \varphi)^2 + \frac{1}{4} \left( 15 e^{ 2 \gamma \varphi} + 10 e^{ - 4 \gamma \varphi} - e^{ - 10 \gamma \varphi} \right) \right) \;,
\end{equation}
where $ \gamma = \sqrt{2/15}$, that emerges as consistent truncation of five--dimensional $ \mathcal{N}=8$ gauged supergravity \cite{Craps2007,Gunaydin1985d}. In this case, the Breitnerlohner--Freedman  bound $m_{BF}^2 = -4$ is saturated:
$$
\varphi = \frac{ \alpha\, \log{r}}{r^2} + \frac{ \beta} {r^2}\;.
$$
Gauged supegravity is thought to be a consistent truncation of ten--dimensional type IIB supergravity so, with usual boundary conditions $ \alpha = 0$, we know the dual theory to be $ \mathcal{N} = 4$ Super--Yang--Mills on $ \mathbb{R} \times S^3$. The bulk scalar $ \varphi$ couples to a trace operator of the scalars in the adjoint:
\begin{equation} \label{eq:dualOperator}
\beta = \langle \mathcal{O}\rangle , \quad\mathcal{O} = c\, \textrm{Tr} \left( \Phi_1^2 - \frac{1}{5} \sum _{j=2}^6 \Phi _{j}^2 \right) \;.
\end{equation}
Imposing generalized $ \alpha = - W_{, \beta}$ boundary conditions corresponds to deforming the dual SYM action \cite{Witten2001}:
\begin{equation}\label{eq:SYMdeformation}
S_{SYM} \rightarrow S_{SYM} - \frac{R_{AdS}^3}{8 \pi G_5} \int dt\, d^3 \theta\, \sqrt{ \Gamma}\, W( \mathcal{O}) \;,
\end{equation}
where $ \Gamma$ is the metric on unit $S^3$.
\begin{figure}[t]
\begin{minipage}{\smallWidthLeft}
\includegraphics[width=\smallWidthRight]{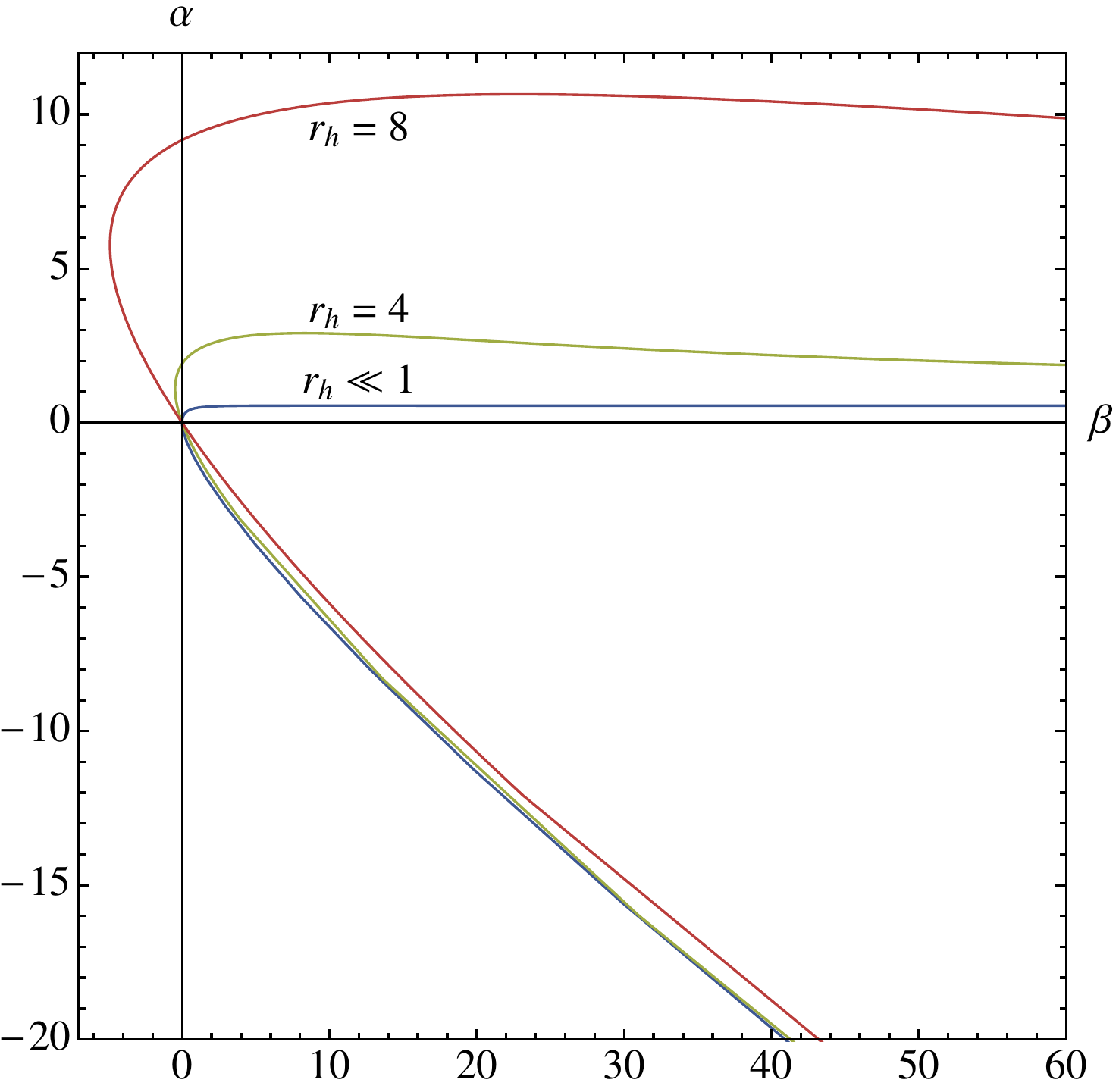}
\end{minipage}%
\begin{minipage}{\smallWidthRight}
\includegraphics[width=\smallWidthRight]{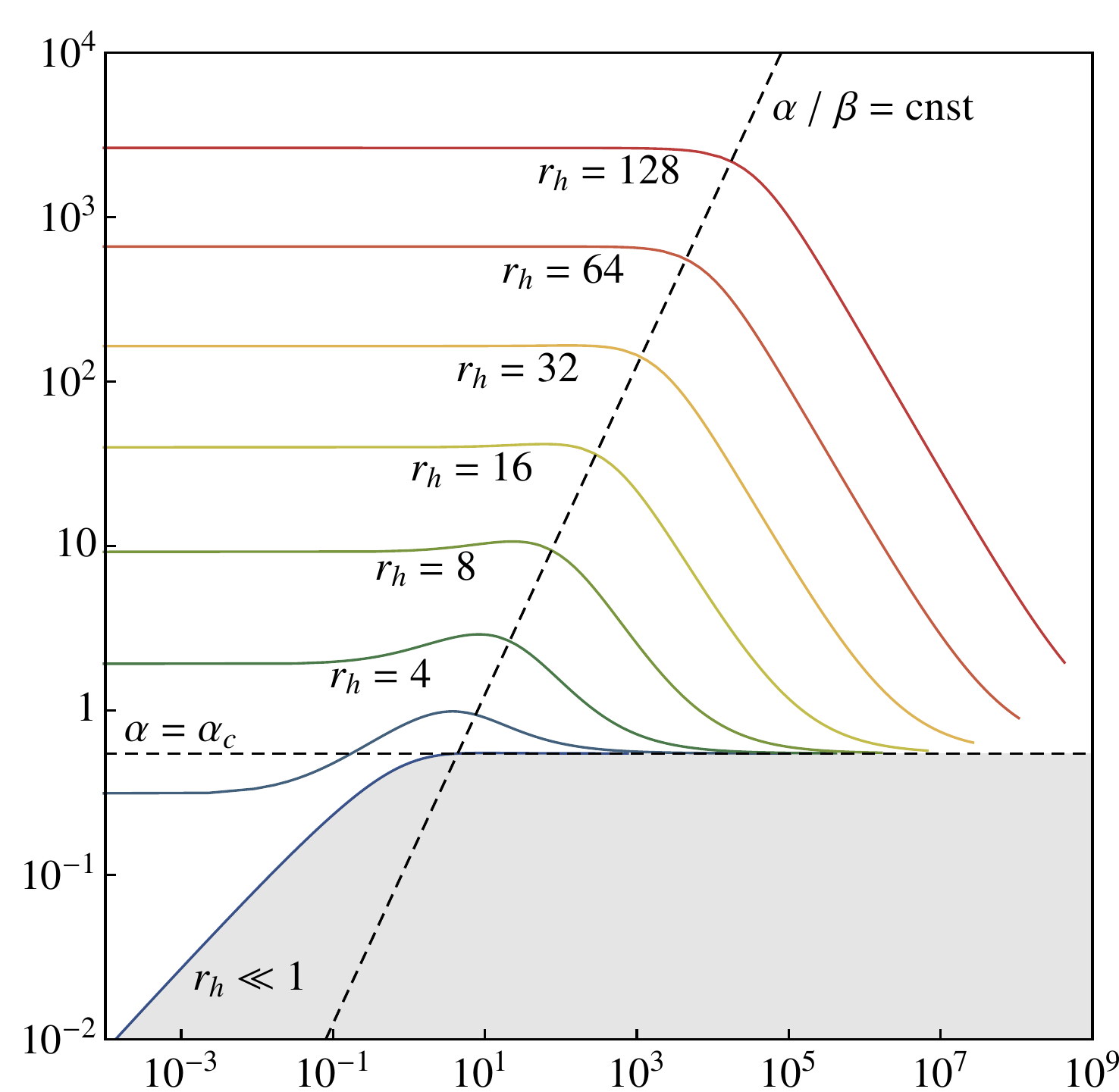}
\end{minipage}
\caption{ \small \label{fig:5DalphaBeta1and2} $ \alpha_{r_h}( \beta)$ curves describing hairy black holes of different radii. In the right panel, log--log plot of the $ \beta > 0$, $ \alpha > 0$ quadrant: all the $ \alpha_{r_h}$ converge to the asymptotic $ \alpha_c$ value with a seemingly $r_h$--independent power law for large $ r_h$. The gray zone contains no black hole solutions.}
\end{figure}%

For positive $ \varphi$, the bulk potential $V$ is concave and one can find numerically a hairy black hole solution for any value of $ r_h$ and $ \varphi_h >0$, with the scalar field vanishing at infinity. At $ \varphi = \varphi_{-} <0$, the potential has a local minimum that corresponds to a smaller radius AdS, and hairy black holes can be found only for $ \varphi_h > \varphi_{-}$: below this value, the potential starts increasing exponentially, and the the resolution of the field equations leads to $ \varphi \rightarrow - \infty$ at a finite value of the radial coordinate.
\begin{figure}[t]
\begin{minipage}{\smallWidthLeft}
\includegraphics[width=\smallWidthRight]{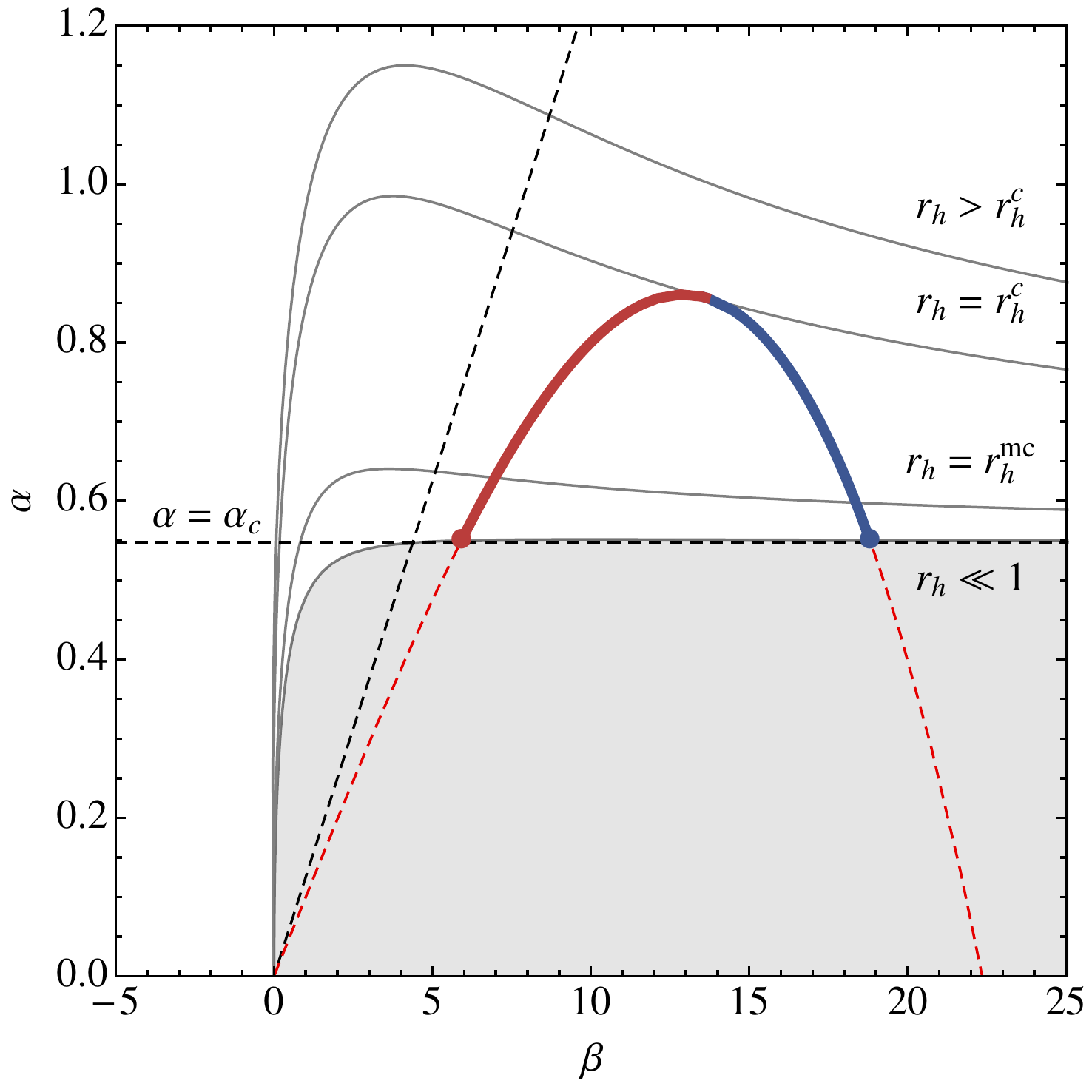}
\end{minipage}%
\begin{minipage}{\smallWidthRight}
\includegraphics[width=\smallWidthRight]{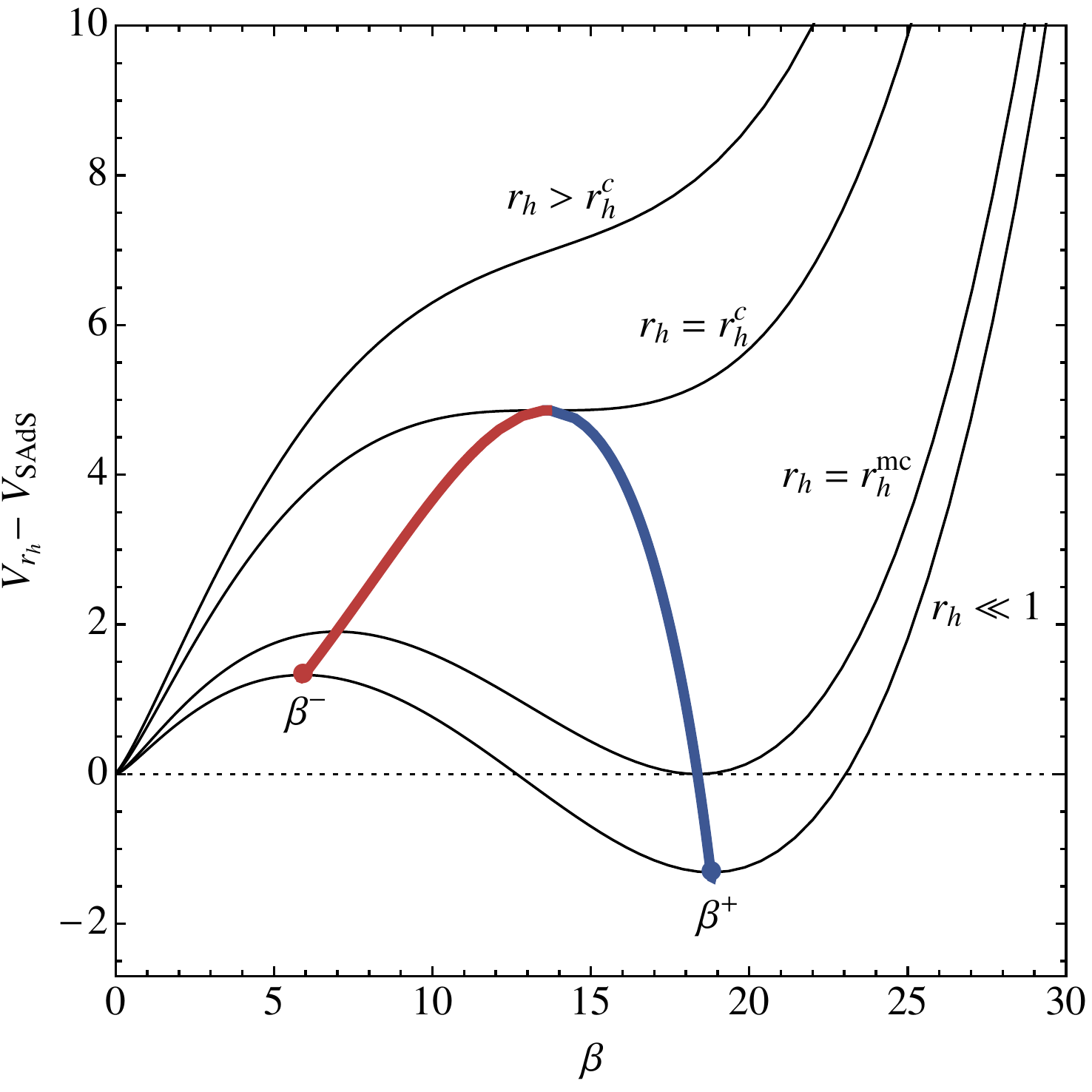}
\end{minipage}
\caption{ \small \label{fig:5Dfamily1and2} In the left panel, intersection of the boundary conditions curve $ \alpha_{f, \eta}( \beta)$ ($f = 10^{-1}$, $ \eta = 0.2$, red dashed curve) with some $ \alpha_{r_h}( \beta)$ curves. Each intersection point corresponds to a single hairy solution compatible with the boundary conditions. For every $r_h \leq r_h^c$, one finds two distinct hairy black holes: the two resulting branches are plotted in thick, solid lines. In the right panel, plot of the corresponding constant entropy effective potentials ($8 \pi G_5 = 1$), whose extrema correspond to the same hairy black hole solutions: at $ r_h = r_h^{c}$, the two non--trivial extrema disappear.}
\end{figure}
Some $ \alpha_{r_h}$ curves are plotted in Figure \ref{fig:5DalphaBeta1and2}: for positive large $ \varphi_h$, $ \alpha_{r_h}$ converges to an asymptotic value $ \alpha_c \simeq 0.54772$, via a power law behavior that seems to be radius--independent at large $r_h$ (right panel). On the contrary, $ \beta$ becomes arbitrarily large. In this region, no black hole is found with $ \alpha \lesssim \alpha _{c}$. When $ r_h \ll 1$, as we have pointed out in the previous section, $ \alpha_{r_h}$ coincides with the $ \alpha_s$ curve described by hairy soliton solutions. At large $ \beta$, $ \alpha_s$ converges to $ \alpha_c$ with a logarithmic error that is small for $ \beta \gtrsim 10$.

For negative values of $ \phi_h$, one finds $ \beta >0$ and $ \alpha < 0$. When $ \phi_h$ approaches $ \phi_-$ from above, both $ \beta$ and $ \alpha$ diverge with a linear behavior corrected by logarithmic terms in $ \beta$: it takes more and more ``time'' (radial coordinate) for the scalar field to escape the fixed point represented by the smaller radius AdS solution, that corresponds to constant negative $ \varphi$, thus to infinite negative $ \alpha$. However, following previous studies, our analysis has focused on the hairy solutions with $ \varphi>0$.

We shall consider the following boundary conditions:
\begin{eqnarray*}
W & = & - \frac{ f}{2} \beta^2 + \frac{ f^3 \eta}{4} \beta^4   \;, \\
\alpha_{f, \eta} & = & f \beta - f^3 \eta \beta^3 \;,
\end{eqnarray*}
where $ f$ and  $\eta$ are positive parameters. Similar boundary conditions were considered, in a very different perspective, in \cite{Battarra2010}: the dynamics at short times of the dual field theory was explored in the limit $ \eta \rightarrow 0$, in which suitable initial conditions evolve into  a cosmological singularity. When $ \eta$ is finite, it was argued that the same evolution ends with the formation of a large hairy black hole \cite{Craps2007}, of the kind we are going to study.

When $ f$ is small, say $ f \lesssim 0.1$, $ \alpha_{f, \eta}$ crosses the soliton curve $ \alpha_s$ at $ \beta \gtrsim 10$ where $ \alpha_s \simeq \alpha_c$. Hence, the effective potential for solitons \eqref{eq:effPotsEq} takes the approximate form:
$$
8 \pi G_5\,\mathcal{V}_s \simeq \alpha _{c} \beta  - \frac{ f}{2} \beta^2 + \frac{ f^3 \eta}{4} \beta^4 \;.
$$
When $ \eta$ is small enough:
$$
\eta \lesssim \eta _{c} = \frac{4}{27 \alpha _{c}^2} \simeq 0.495 \;,
$$
$ \mathcal{V}_s$ has two non trivial extrema $ \beta _{ \pm}$ that correspond to hairy soliton solutions compatible with the required boundary conditions (see Figure \ref{fig:5Dfamily1and2}). The less hairy soliton, located at local maximum $ \beta_{-}$  of $ \mathcal{V}_s$, has positive mass. The more hairy one, located at the bottom $ \beta _{+}$ of $ \mathcal{V}_s$ has negative mass provided $ \eta \lesssim \eta_c/2$. Two distinct branches of hairy black holes start from these solitons, and end by joining to each other at some finite radius $r_h = r_h ^{c}$. Accordingly, while $r_h$ grows, the shape constant entropy effective potential changes continuously and, for $r_h > r_h^{c}$ no more extrema are present. In between this critical point and the solitons, above some intermediate value $ r_h^{mc}$, the more hairy black hole becomes more massive than the Schwarzschild--AdS (SAdS) solution of the same radius.

\subsection*{Quasi--normal frequencies}
The shape of the soliton effective potential suggests that the less hairy soliton may be unstable under perturbations of the scalar field. Such instability has indeed been found in \cite{Hertog2005b} with simpler boundary conditions:
$$
\alpha = \alpha_{f, 0} \equiv f \beta \;,
$$
that give rise to a soliton potential that is unbounded from below. The authors have found that the corresponding black holes possess a growing mode with pure imaginary frequency $ \omega = i \nu_0$, $ \nu_0 >0$ (see \eqref{eq:linearAnsatz1}). 

One can actually prove that, if such instability exists, the associated frequency ought to be pure imaginary. To simplify the notation, let us choose $r ^{*}(r = \infty) = 0$. Multiplying \eqref{eq:perturbations} by $ \Phi ^{*}$ and integrating between $r ^{*} = - \infty$, corresponding to the horizon, and $r ^{*} = - \epsilon$, one finds:
$$
\int_{- \infty} ^{ - \epsilon} d r ^{*}\left(  \Phi ^{\star} \frac{d^2 \Phi}{d r ^{* 2}} - f _0 e^{-2 \delta_0} V_0 | \Phi|^2 + \omega^2 | \Phi|^2 \right)  = 0 \;.
$$
For a growing mode of frequency $ \omega = \omega_R + i \omega_I$, $ \omega_I >0$, ingoing boundary conditions at the horizon:
$$
\Phi \sim e^{- i \omega r ^{*}} \;,
$$
imply that $ \Phi$ decreases exponentially when $ r ^{*} \rightarrow - \infty$, so the integral above is perfectly finite. Taking the imaginary part and integrating by parts, we obtain:
$$
\textrm{ Im} \left( \Phi ^{\star} \frac{d \Phi}{d r ^{*}} \right) (r ^{*} =  - \epsilon) + 2\, \omega_R\, \omega_I \int_{- \infty} ^{ - \epsilon} d r ^{*}\, |\Phi|^2 = 0 \;.
$$
The coefficient function $ f_0 e^{-2 \delta_0} V_0$ has, near the AdS boundary, the following behavior:
$$
f_0 e^{-2 \delta_0} V_0 = - \frac{1}{4} |r ^{*}|^{ -2} + \mathcal{O}(1) \;.
$$
Hence, asymptotically:
$$
\Phi = - A\, |r ^{*}|^{ 1/2}\, \log{|r ^{*}|} + B\,   |r ^{*}|^{ 1/2}  + \mathcal{O} \left( \log{ |r ^{*}|} \,|r ^{*}|^{ 5/2} \right) \;.
$$
Imposing our boundary conditions \eqref{eq:linBcs2} requires $A$ and $B$ to have the same phase, so near the boundary:
\begin{equation}\label{eq:proof}
\textrm{Im} \left( \Phi ^{\star} \frac{ d \Phi}{d r ^{*}} \right) = \mathcal{O} \left( \log{ | r ^{*}|}\, | r ^{*}|^2 \right) \;.
\end{equation}
So, moving to the limit $ \epsilon \rightarrow 0 ^{+}$ we deduce $ \omega_R = 0$.

\begin{figure}[t]
\begin{minipage}{\smallWidthLeft}
\includegraphics[width=\smallWidthRight]{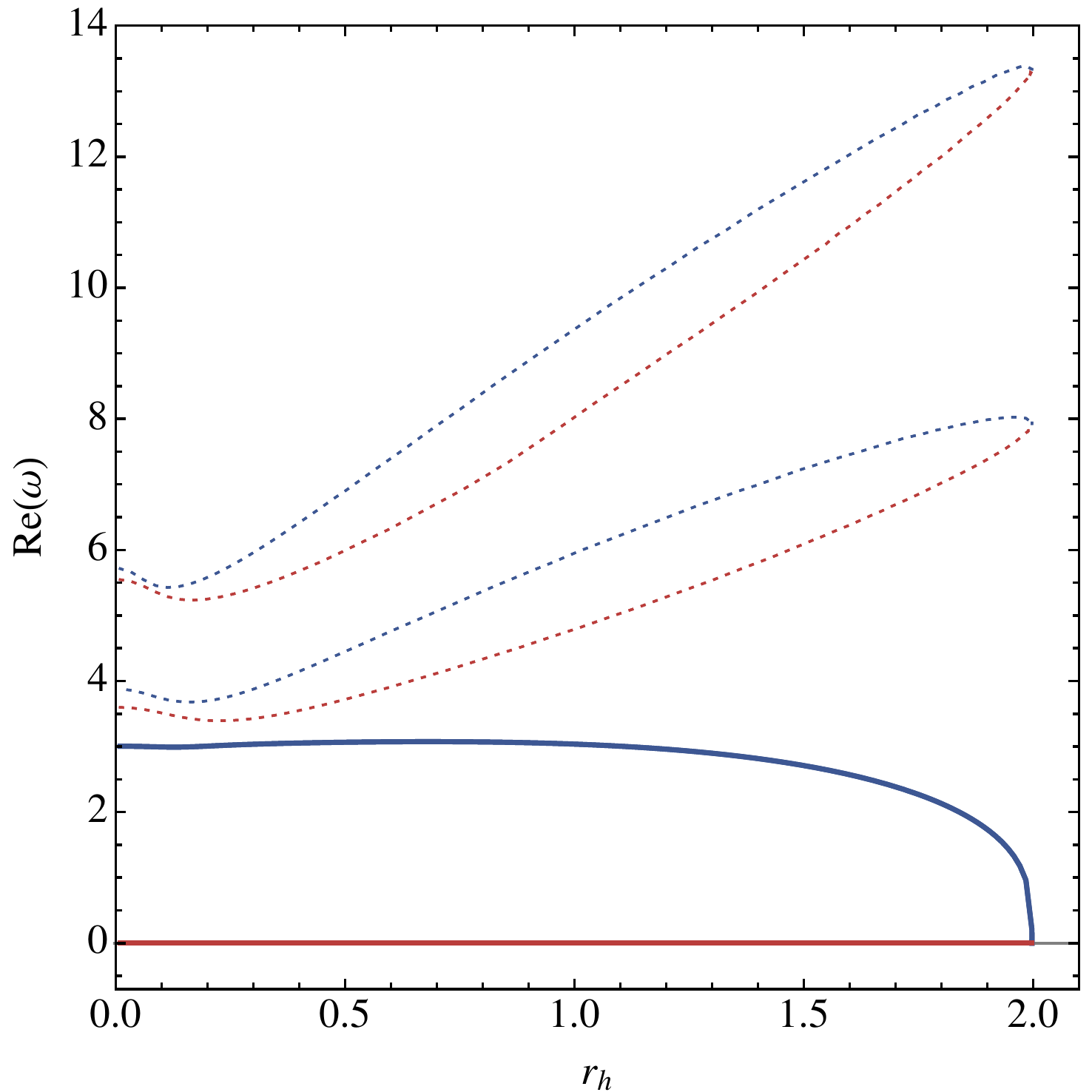}
\end{minipage}%
\begin{minipage}{\smallWidthRight}
\includegraphics[width=\smallWidthRight]{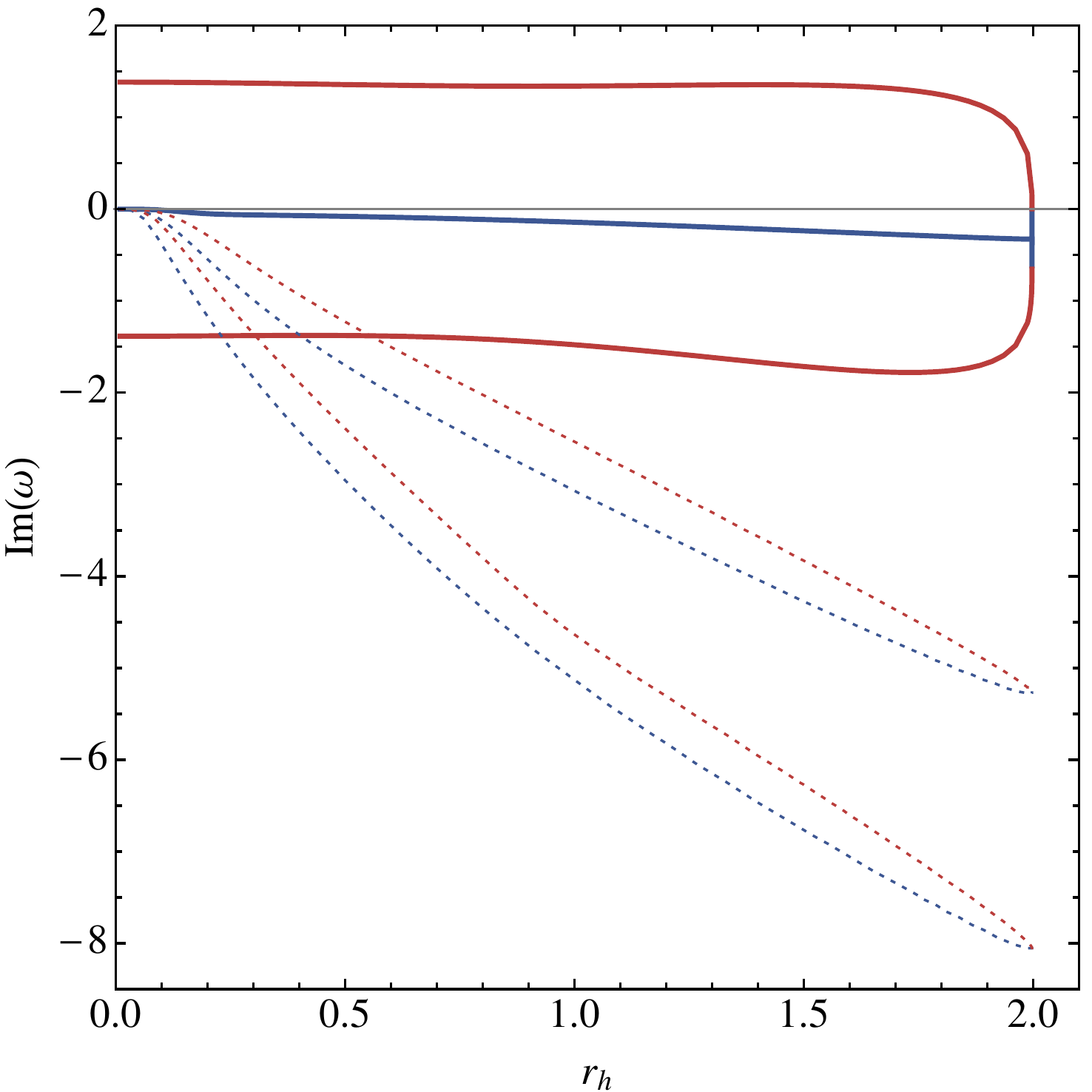}
\end{minipage}
\caption{ \small \label{fig:5DfamilyModes1and2} $ f=0.1$, $ \eta = 0.2$. Real and imaginary part of $ \omega_0$, $ i \nu_0$, $ i \tilde{\nu}_0$ (solid lines, same colors as in Figure \ref{fig:5Dfamily1and2}) and further quasi--normal frequencies $ \omega_1$ and $ \omega_2$ (dotted lines). Positive imaginary part is associated with growing modes. At the critical point $r_h = r_h^c$, $ \omega_0$ and $ i \nu_0$ vanish. For a more detailed plot of the critical point $r_h = r_h^c$, see Figure \ref{fig:5DfamilyModes3and4}.}
\end{figure}%

Going back to our $ \alpha_{f, \eta}$ boundary conditions, we observe that when $ \eta$ is small, the shape of the effective potential around its unstable maximum is essentially $ \eta$--independent, so we expect to find the same instability: our numerical data confirms that a growing mode characterizes the less hairy branch of black holes with $ \alpha_{f, \eta}$ boundary conditions as well. This frequency is accompanied by another pure imaginary but damped frequency $ \omega = i \tilde{ \nu}_0$, $ \tilde{ \nu}_0 < 0$. In particular, we find $ \nu_{0}(r_h = 0) = - \tilde{ \nu}_0(r_h = 0)$, and this comes as no surprise: a given regular solution $ \Phi_{ \omega}$ of \eqref{eq:perturbations} solves both for $ \omega$ and $ - \omega$, translating the symmetry of both the wave equation and the boundary conditions under time reversal. At finite radius $ r_h >0$,  an ingoing solution $ \Phi_ \omega$ still solves for $ - \omega$ too, but it is purely outgoing with respect to this frequency\footnote{On the contrary, $ \Phi _{\omega} ^{\star}$ is an ingoing solution for $ - \omega ^{*}$, and the quasi--normal frequencies come as usual in couples $( \omega, - \omega ^{*})$.}. Accordingly, we find $ \nu_0 \neq -\tilde{ \nu}_0$ at finite radius.

\begin{figure}[t]
\begin{minipage}{\smallWidthLeft}
\includegraphics[width=\smallWidthRight]{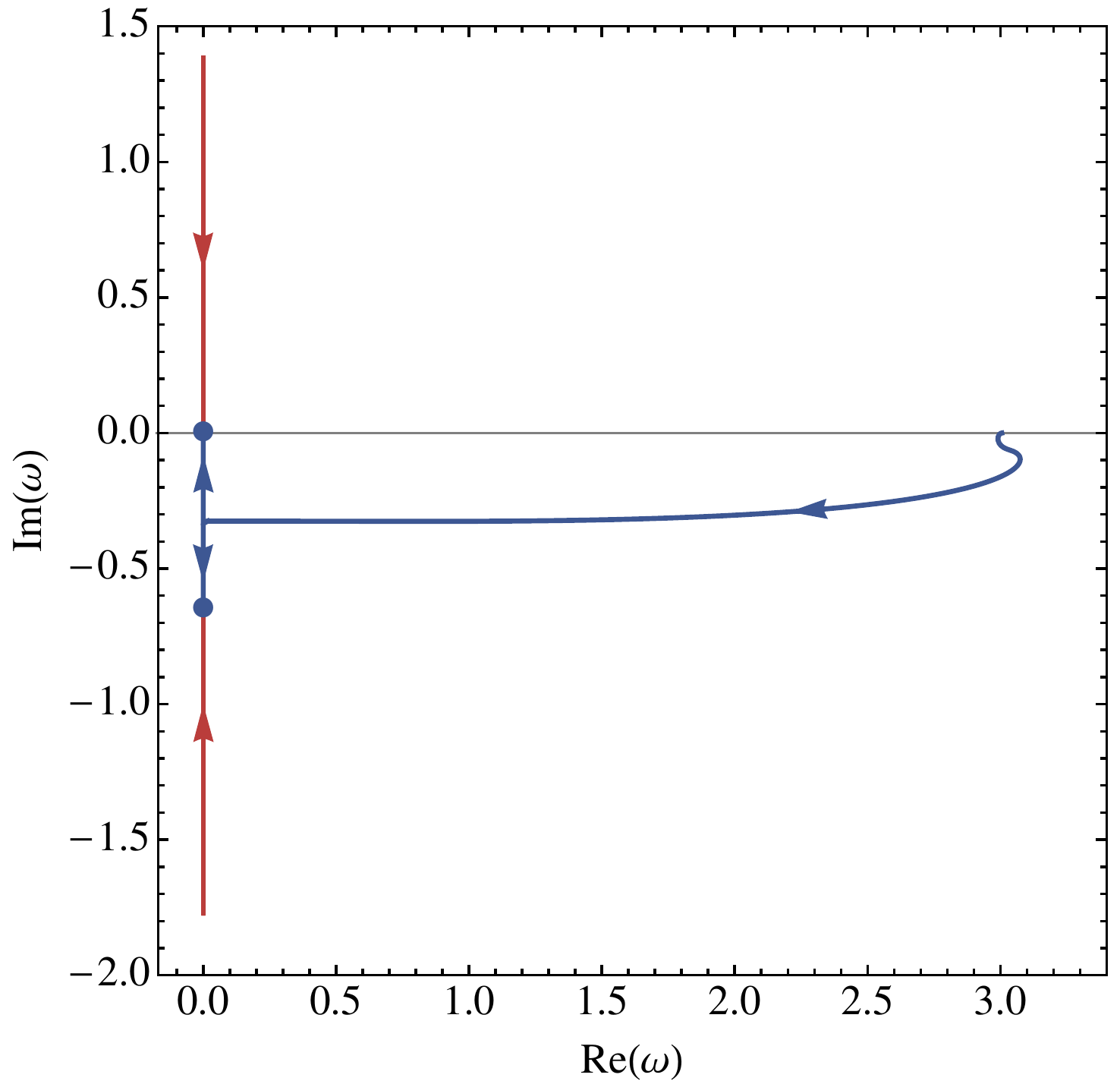}
\end{minipage}%
\begin{minipage}{\smallWidthRight}
\includegraphics[width=\smallWidthRight]{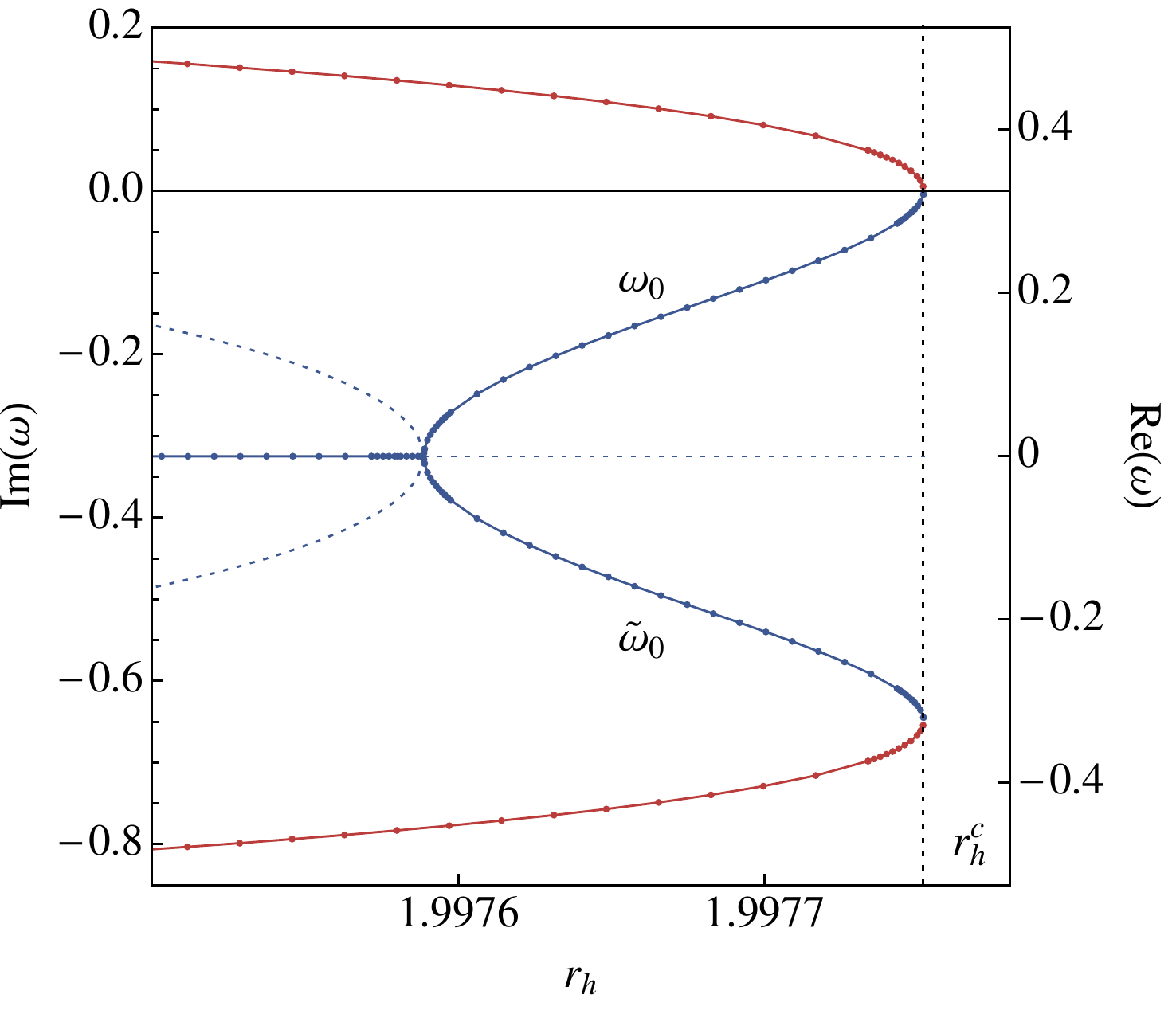}
\end{minipage}
\caption{ \small \label{fig:5DfamilyModes3and4} In the left panel, trajectory of $ \omega_0$, $ \tilde{ \omega}_0$, $ i \nu_0$ and $ i \tilde{ \nu}_0$ in the complex plane; arrows correspond to the direction of increasing radius, and the blue points mark the critical radius $ r_h^c$, where $ \omega_0 = \nu_0 = 0$. In the right panel, a detailed plot of $ \textrm{Im}( \omega_0)$ and $ \textrm{Im}( \tilde{ \omega}_0)$ near $ r_h = r_h^c$; $ \textrm{Re}( \omega_0)$ is plotted in dotted lines. At the bifurcation point, the $ \omega_0$ frequency characterizing the stable branch becomes pure imaginary and gives rise to two distinct pure imaginary frequencies $ \omega_0$ and $ \tilde{ \omega}_0$, one of which turns to zero frequency when $ r_h \rightarrow r_h^{c}$.}
\end{figure}%

When the radius is increased and approaches the critical value $ r_h^{c}$, $ \nu_0( r_h)$ approaches zero (see Figure \ref{fig:5DfamilyModes1and2}). Indeed, at $ r = r_h^{c}$, the boundary condition curve $ \alpha_{f, \eta}$ and $ \alpha_{r_h^{c}}$ are tangent to each other, and the static perturbation giving rise to the infinitesimal static displacement along this tangent is compatible with the boundary conditions. On the contrary, the spurious frequency $ i \tilde{ \nu}_0( r_h)$ stays away from zero at $ r=r_h^{c}$. These frequencies can be followed back along the branch of more hairy black holes, that starts at $r_h = 0$ from the soliton that corresponds to the global minimum of $ \mathcal{V}_s$. We will denote $ \omega_0( r_h)$ and $ \tilde{ \omega_0}(r_h)$ the related frequencies (see Figure \ref{fig:5DfamilyModes3and4}):
\begin{eqnarray*}
\omega_0( r_h^c) & = & i \nu_0(r_h^c) = 0 \;,\\
\tilde{ \omega}_0(r_h^c) & = & i \tilde{ \nu}_0(r_h^c) \;.
\end{eqnarray*}
For values of the radius extremely close but smaller than $r_h^{c}$, $ \omega_0$ and $ \tilde{ \omega}_0$ are still pure imaginary, but $ \omega_0$ has now a negative imaginary part corresponding to damped oscillations. Within a typically tiny distance from $ r_h^{c}$, $ \omega_0$ joins with $ \tilde{ \omega}_0$ (see Figure \ref{fig:5DfamilyModes1and2}) at some point of the imaginary axis. For smaller values of $ r_h$, the remaining frequency acquires an oscillating part $ \textrm{Re}( \omega_0) \neq 0$ and is now distinct from his partner at $ \omega = - \omega_0 ^{\star}$. Finally it ends, at $ r_h = 0$, in a pure real normal frequency of the more hairy soliton. A very similar bifurcation has been observed in the analysis of vector quasi--normal frequencies of D7 branes \cite{Kaminski2010a}, with momentum replacing $ r_h$ as driving parameter. In that context, the presence of a $ \omega = 0$ mode at $ k=0$ translates the homogenous diffusion of a conserved charge in the presence of translational symmetry.

In summary, with $ \alpha_{f, \eta}$ boundary conditions, the qualitative picture given by the evolution of the constant radius effective potential (see Figure \ref{fig:5Dfamily1and2}) turns out to be accurate in predicting the perturbative stability/instability of the hairy black holes under perturbations of the hair itself.

Apart from the one we have just described, both branches of black holes possess a presumably infinite spectrum of other quasi--normal frequencies. They all correspond to damped perturbations and, as we have already pointed out, they are characterized by pure real frequencies in the limit $r_h \rightarrow 0$. By continuity between the stable and unstable branches, their trajectory as a function of the radius forms closed curves in the complex plane  (see Figure \ref{fig:5DfamilyModes1and2}). In the range of parameters we have explored, the sequence of these ``standard'' quasi--normal frequencies respects the order one finds, for example, for standard perturbations of SAdS black holes: they can be labeled in such a way that both $ \textrm{Re}( \omega_n)$ and $ | \textrm{Im}( \omega_n)|$ increase with $ n$ and this, for any $ r_h$. On the stable branch,  $\omega_0$ does not always respect such order. In fact $ \omega_0$ turns out to be the longest lived frequency but, for small values of $ \eta$, its real part can be greater than some of the $ \textrm{Re}( \omega_n)$.

\subsection*{Parameter dependence and effective potentials}
Let's consider again the approximate expression of the soliton effective potential in the $ \beta \gtrsim 10$ region:
\begin{equation} \label{eq:solitonEffPot}
\mathcal{V}_s \simeq \frac{\alpha_c \beta + W( \beta)}{8 \pi G_5} = \frac{N^2}{4 \pi^2} (\alpha_c \beta + W( \beta)) \;,
\end{equation}
where we have made use of the standard AdS/CFT relation:
$$
\frac{R_{AdS}^3}{8 \pi G_5} = \frac{N^2}{4 \pi^2} \;.
$$
Recalling the form of the dual operator \eqref{eq:dualOperator}, we observe that the linear term in $ \mathcal{V}_s( \beta)$ takes the form of a mass term for the scalars. Such kind of term is already present in the undeformed SYM action as conformal coupling of the scalars to the curvature of $S^3$:
\begin{equation} \label{eq:SYMpot}
S_{SYM} = \int dt\, d^{3} \theta \sqrt{ \Gamma} \left[ \textrm{Tr} \left(- \frac{1}{2} D _{\mu} \Phi _{i} D ^{\mu} \Phi _{i}- \frac{1}{2} \Phi _{i}^2 \right) - \frac{ W( \mathcal{O})}{8 \pi G_5} \right] + \ldots \;.
\end{equation}
Moreover, a classical analysis shows that the extrema of the SYM  deformed scalar potential, that may be identified with our hairy solitons, are characterized by $ \Phi_i = 0$ for $i=2,\ldots,6$. Therefore, comparing \eqref{eq:solitonEffPot}, \eqref{eq:SYMpot} and \eqref{eq:dualOperator}, we tentatively identify:
\begin{equation} \label{eq:operatorNormalization}
c = \frac{2 \pi^2}{\alpha_c N^2} \;.
\end{equation}
At this stage, this is a simple prescription for the soliton effective potential to coincide with the deformation plus a free term in the original SYM action. The masses of the hairy solitons can now be obtained as vev of the SYM scalar potential, integrated over the volume of $ S^3$. However, such prescription can be checked by an independent calculation, e.g. matching the normalization of the correlation function of $ \mathcal{O}$ in the field theory.  As we shall show in the next section, the mere existence of such prescription is not guaranteed, but is rather a property of the particular consistent truncation we are considering: in fact, the key point is that $ \alpha_s \rightarrow \alpha_c$ when $ \beta \rightarrow \infty$. It should also be stressed that, as the authors of \cite{Hertog2005b} have pointed out in a four dimensional case, the mass term does not appear at small $ \beta$, where $ \alpha_s$ grows as $ \beta^{1/2}$.

We can cross--check our prescription via the quasi--normal frequencies data. We introduce a further simplification by tentatively neglecting the non--abelian structure of the Higgs scalar $ \Phi_1$:
\begin{eqnarray}
\Phi_1 & = & \sqrt{N} \phi\, I_N \;,\\ \label{eq:betaNormalization}
\beta & = &  \frac{2 \pi^2}{ \alpha_c} \phi^2 \;.
\end{eqnarray}
Neglecting all the other fields in \eqref{eq:SYMpot}, we consider the following toy model free action for the dual theory:
\begin{eqnarray}
\label{eq:toyAction}S & = & N^2 \int dt\, d^{3} \theta \sqrt{ \Gamma} \left( - \frac{1}{2} ( \partial _{\mu} \phi)^2- \frac{1}{2} \phi^2 - \frac{1}{4 \pi^2} W_{f, \eta} \left( \frac{2 \pi^2}{ \alpha_c} \phi^2 \right) \right)\\ \nonumber
& = & N^2 \int dt \,d^3 \theta\, \sqrt{ \Gamma}\left(- \frac{1}{2} ( \partial \phi)^2 - \frac{ \mathcal{V}_s}{N^2} \right) \;.
\end{eqnarray}
At the extrema of $ \mathcal{V}_s$, the mass of the scalar field $ \phi$ is given by:
\begin{equation} \label{eq:frequencyProposal}
m_s^2 \simeq 4\, c\, \beta \frac{d^2 \mathcal{V}_s}{ d \beta^2} = \frac{2 \beta}{ \alpha_c} \left( \frac{d \alpha_{s}}{d \beta} + W_{, \beta \beta} \right) \;.
\end{equation}
\begin{figure}[t]
\begin{minipage}{\smallWidthLeft}
\includegraphics[width=\smallWidthRight]{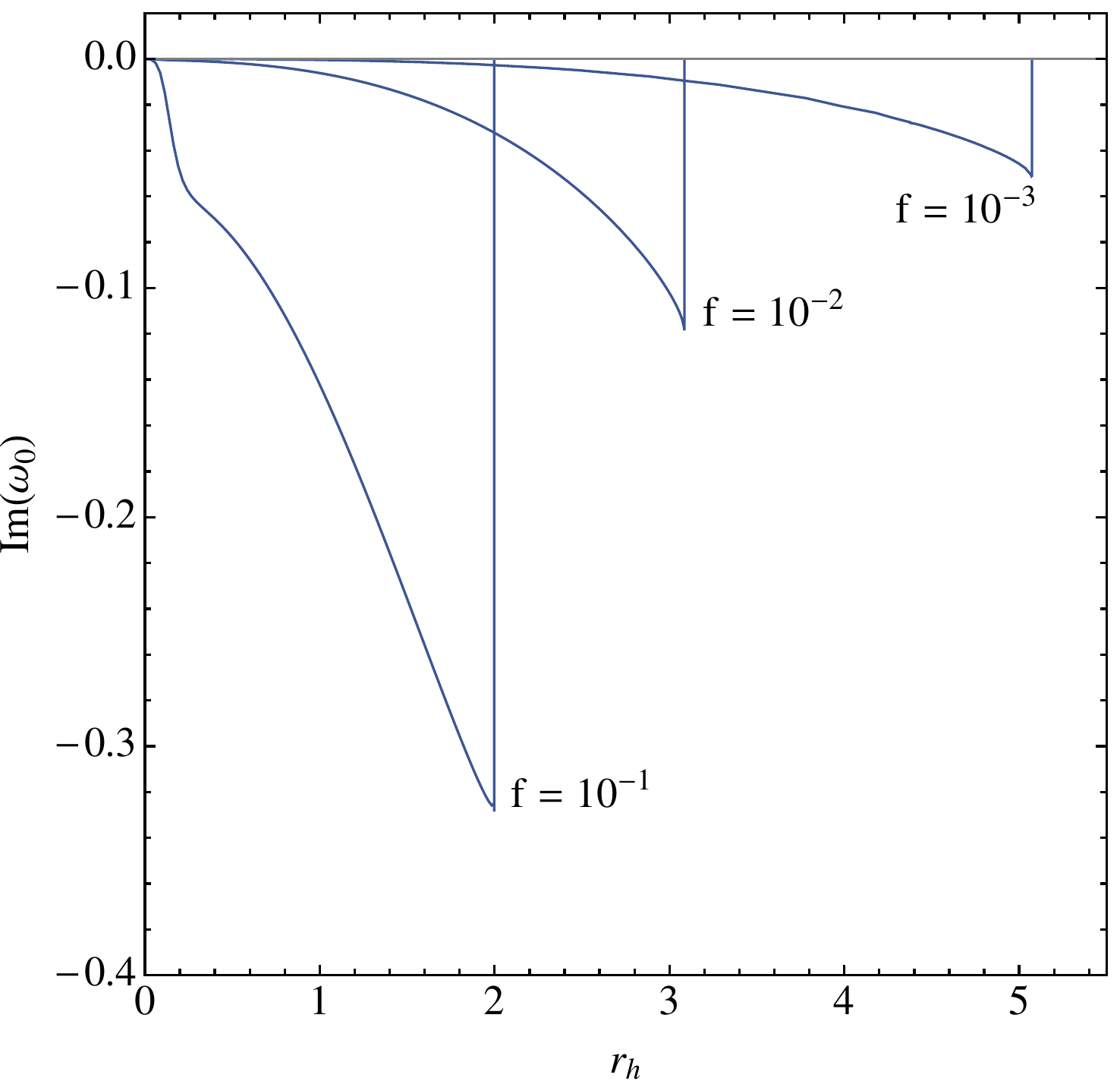}
\end{minipage}%
\begin{minipage}{\smallWidthRight}
\includegraphics[width=\smallWidthRight]{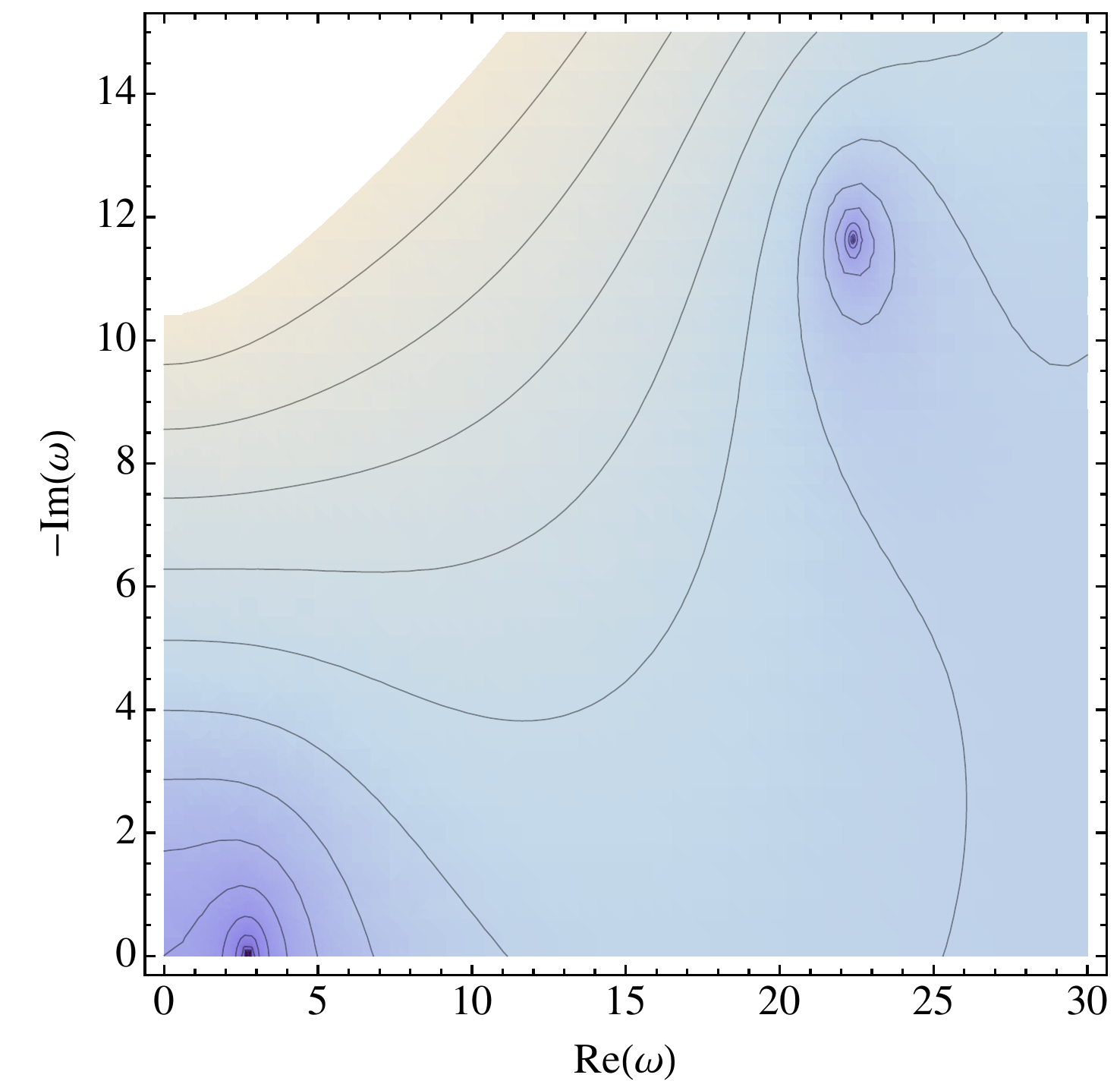}
\end{minipage}
\caption{ \small \label{fig:5Dconjecture1and2} In the left panel, imaginary part of $ \omega_0$ as a function of the radius, for $ \eta = 0.2$ and $ f = 10^{-1}$,$10^{-2}$ and $10^{-3}$. The cusps coincide with the bifurcation point (see Figure \ref{fig:5DfamilyModes3and4}). When $f \rightarrow 0$, corresponding to the large field limit, the $ \omega_0$ becomes stable. In the right panel, the goal function whose roots coincide with the quasi--normal frequencies, for the hairy black hole of radius $ r_h \simeq 2.4$ and compatible with boundary conditions $ \eta = 0.2$, $ f = 10^{-3}$: the $ \omega_0$ frequency has an imaginary part of order $ - 10^{-2}$, while the second overtone $ \omega_1$ has $ \textrm{Im}( \omega_1) \simeq - 12$.}
\end{figure}%
In particular:
\begin{eqnarray*}
m_s^{2}( \beta _{-}) & \equiv  &- \nu_s^2 < 0 \;,\\
m_s^2( \beta_{+}) & \equiv & \omega_s^2 > 0 \;.
\end{eqnarray*}
We propose to compare these masses to the values of $-\nu_0^2$ and $ \omega_0^2$ that characterize the unstable/stable solitons respectively. We  generalize this proposal to match the values of $ \nu_0^2$ and $ \omega_0^2$ for $ r_h >0$: we already know that, when $ r_h$ approaches $ r_h^{c}$, one finds $ \nu_0, \omega_0 \rightarrow 0$. At the same time, the curvature of the effective potential $ \mathcal{V}_{r_h}$ near its unstable maximum vanishes (see Figure \ref{fig:5Dfamily1and2}). This suggests the minimal generalization of $ \eqref{eq:frequencyProposal}$ that consists in replacing $ \mathcal{V}_s$ with $ \mathcal{V}_{r_h}$:
$$
m_{r_h}^2 \equiv 4\, c\, \beta \frac{d^2 \mathcal{V}_{r_h}}{ d \beta^2} = \frac{2 \beta}{ \alpha_c} \left( \frac{d \alpha_{r_h}}{d \beta} + W_{, \beta \beta} \right) \;,
$$
and accordingly:
\begin{eqnarray*}
m_{r_h}^2( \beta _{-}) & \equiv & - \nu_{r_h}^2 < 0 \;,\\
m_{r_h}^2( \beta _{+}) & \equiv & \omega_{r_h}^2 > 0 \;.
\end{eqnarray*}
The derivative $ d \alpha_{r_h}/ d \beta$ can be easily evaluated numerically. Of course, we cannot hope to describe in this simple way the dissipation described by the imaginary part of $ \omega_0$ on the stable branch of hairy black holes: $ \omega_{r_h}^2$ is real while $ \omega_0^2$ is not. Therefore, we limit ourselves to comparing $ \omega_{r_h}^2$ with the real part of $ \omega_0^2$ as a function of the horizon radius.

\begin{figure}[t!]
\includegraphics[width=\hugeWidth]{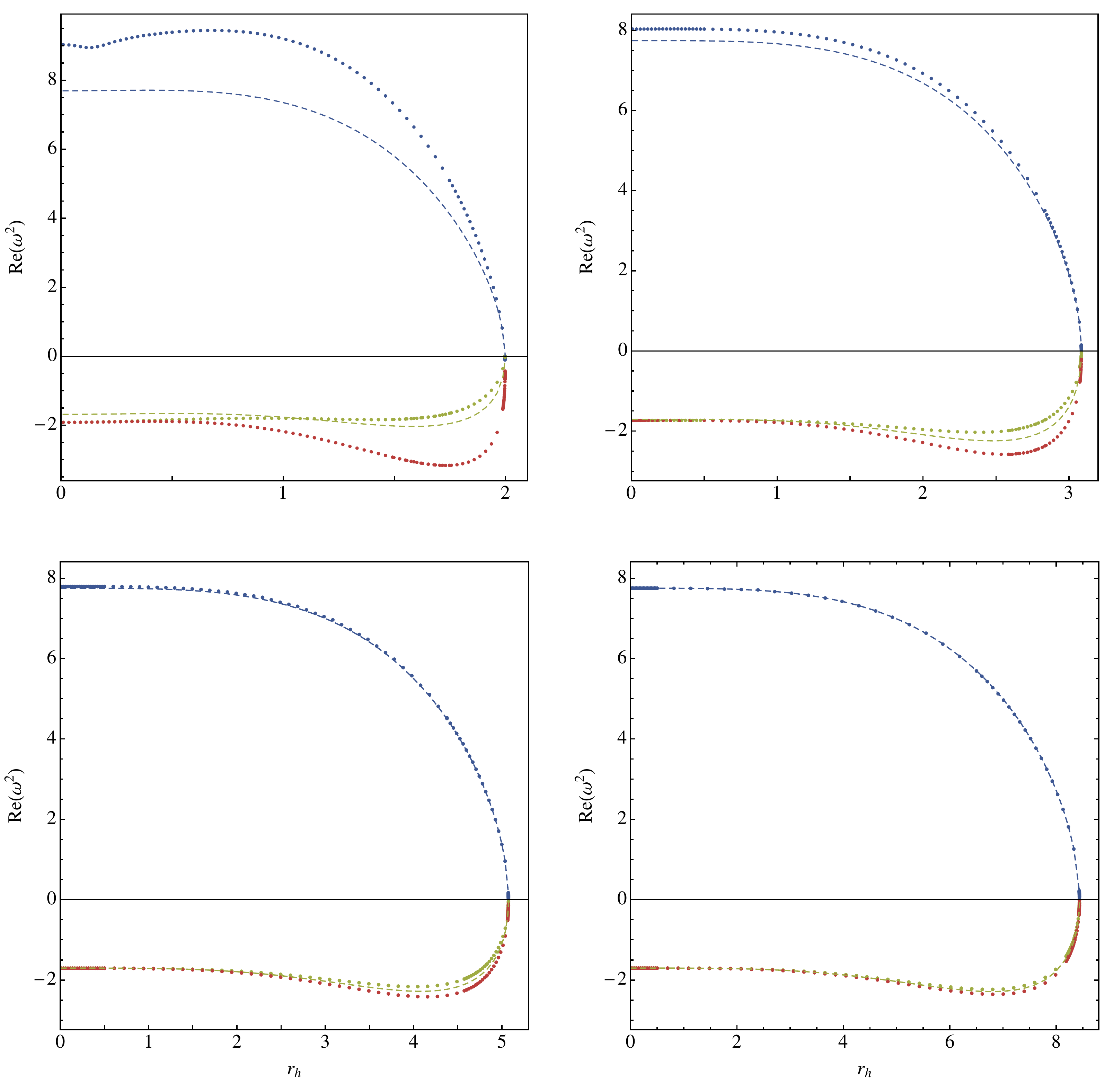}
\caption{ \small \label{fig:5Dconjecture} Comparison between the real part of $ \omega_0^2$ and the curvature of the constant entropy effective potential at its extrema, for boundary conditions with $ \eta = 0.2$ and $f=10^{-1}$ (top left panel), $f=10^{-2}$ (top right), $f=10^{-3}$ (bottom left), $f = 10^{-4}$ (bottom right). Scattered points are quasi--normal frequencies $ \omega_0$ (stable branch) and $ i \nu_0$, while green dots represents the damped, pure imaginary frequencies $ i \tilde{ \omega}_0$ characterizing the unstable black holes. Dashed lines represent the second derivative of the radius--dependent constant entropy effective potential, evaluated at its two extrema. In the large field limit $ f \rightarrow 0$, the agreement becomes better and better, and $ \tilde{ \nu}_0^2 = \nu_0^2$. Similar results have been obtained for smaller values of $ \eta$.}
\end{figure}%
The numerical agreement depends on the values of $f$ and $ \eta$, and it can be better analyzed by taking a closer look at the parameter dependence of the toy model soliton potential for $ \phi$ that appears in \eqref{eq:toyAction}:
$$
V_{f, \eta}( \phi) = \frac{1}{2} \phi^2 - \frac{\tilde{f}}{2} \phi^4 + \frac{\tilde{f}^3 \tilde{ \eta}}{4} \phi^8 \;,
$$
where $ \tilde{f}$ and $ \tilde{\eta}$ include the order one normalization factors in \eqref{eq:toyAction}. Of course, $ m_s^2$ is simply given by the second derivative of $V_{f, \eta}$ with respect to $ \phi$, evaluated at his extrema. The way we parametrized the boundary conditions lets one reabsorb $f$ in a rescaling of $ \phi$ and $V_{f, \eta}$:
\begin{equation} \label{eq:scalingf}
V_{f, \eta}( \phi) = f ^{-1}\, V_{1, \eta} ( \sqrt{f} \phi) \;.
\end{equation}
Thus, when $ f$ decreases, the masses of the stable/unstable solitons and the corresponding values of $ \phi^2_{\pm} \propto \beta _{\pm}$ scale as $ f^{-1}$. At the same time, as one can  deduce from Figure \ref{fig:5DalphaBeta1and2}, the critical radius $ r_h^{c}$ increases; numerical data shows that:
\begin{equation} \label{eq:rhEndScaling}
r_{h}^{c} \propto f^{-0.22} \;.
\end{equation}
From \eqref{eq:scalingf}, one sees that $ \omega_s^{2}$ and $ \nu_s^2$ do not scale with $f$:
$$
m_s^2 = \frac{d^2 V_{f, \eta}}{d \phi^2}( \phi _{\pm}) = V_{1, \eta}''( \sqrt{f} \phi _{\pm}) \;.
$$
Therefore, as long as this correspondence holds, $ \omega_0^2(r_h = 0)$ and $ \nu_0^2(r_h=0)$ should reach a finite limit when $ f \rightarrow 0$. This is confirmed by numerical results (see Figure \ref{fig:5Dconjecture}), and also holds for hairy black holes, provided one rescales $r_h$ according to \eqref{eq:rhEndScaling}. Moreover, in the same limit $ f \rightarrow 0$, we observe that that the imaginary part of $ \omega_0$ rapidly goes to zero along the whole branch of stable hairy black holes (see figure \ref{fig:5Dconjecture1and2}). One finds $| \textrm{Im}( \omega_0)| \propto f^{0.3}$ at constant at constant radius, $| \textrm{Im}( \omega_0)| \propto f^{2}$ at constant temperature. This justifies the free character of our toy model action, so as to study the $ \omega_0$ frequency. In a related way, the spurious frequency $ i \tilde{ \nu}_0$ characterizing the unstable black holes appears to coincide with the opposite of its companion $ i \nu_0$, signaling again the approach to a non dissipative regime:
$$
\tilde{ \nu}_0 \stackrel{ f \rightarrow 0}{ \simeq} - \nu_0 \;.
$$
Finally, the values of the curvature of the constant effective potential near its extrema, $ \omega_{r_h}^2$ and $ - \nu_{r_h}^2$, give in this limit a very good estimation of $ \textrm{Re}(\omega_{0}^2)$ and $- \nu_0^2$ (see Figure \ref{fig:5Dconjecture}).

In summary, $ f \rightarrow 0$ corresponds to increasingly large values of the condensate $ \beta$. This limit preserves the curvature of the deformation in terms of boundary scalar $ \phi$: the masses of the solitons scale as $ f^{-1}$, but so does $ \phi^2$. We have reviewed the presence of a particular frequency $ i\nu_0$ (resp. $ \omega_0$), in the spectrum of quasi--normal frequencies of the corresponding unstable (resp. stable) hairy black holes, and showed that its value is very accurately predicted, in the limit $ f \rightarrow 0$, by the value of the curvature of the constant entropy effective potential at its maximum (resp. minimum). More specifically, $ i \nu_0$ is the frequency of the growing mode that is responsible for an instability of the related hairy black holes. This agreement is realized by assuming that the mass term for $ \beta$ appearing in $ \mathcal{V}_s$ is the conformal mass term coupling $ \Phi_1$ to the curvature of $S^3$: under this hypothesis, the normalization of $ \mathcal{O}$, that enters in the expression of $m_{r_h}^2$, is fixed by the numerical value of $ \alpha_c$.  In the same $ f \rightarrow 0$ limit, the spurious frequency $ i \tilde{ \nu}_0$ approaches the value $ - i \nu_0$: therefore, these two frequencies appear as solutions to the toy model equation for homogeneous perturbations:
$$
\delta\ddot{ \phi} + \frac{1}{N^2} \left.\frac{d^2 \mathcal{V}_{r_h}}{d \phi^2}\right|_{ \phi_{-}} \delta\phi = 0 \;,
$$
that generalizes \eqref{eq:toyAction} to finite radius. At the same time, the imaginary part of $ \omega_0$ goes to zero, and $ \omega_0$ solves the analogous equation:
$$
\delta\ddot{ \phi} + \frac{1}{N^2} \left.\frac{d^2 \mathcal{V}_{r_h}}{d \phi^2}\right|_{ \phi_{+}} \delta\phi = 0 \;.
$$
While $ \omega_0$ is longer and longer lived when $ f$ decreases, the remaining ``standard'' quasi--normal frequencies do not share this feature: for example, as we show in figure \ref{fig:5Dconjecture1and2}, the imaginary part of $ \omega_1$ reaches larger and larger negative values, as one usually expects for increasing values of $ r_h$.

In conclusion, the normalization \eqref{eq:operatorNormalization} is quantitatively confirmed by the agreement of the quasi--normal frequencies $ \omega_0, i \nu_0$ with the curvature of the constant radius effective potentials, in the limit of large condensates $ f \rightarrow 0$.  Therefore, we expect $ \alpha_s \rightarrow \alpha_c \neq 0$ to be a consequence of the curvature of the spatial sections of the boundary. In the dual field theory, homogenous excitations of the condensate operator around the minimum of the effective potential are longer and longer lived, due to the presence of a single quasi--stable resonance.  Remarkably, higher overtones do not share this property: their imaginary part remains finite when $f \rightarrow 0$.

\section{Hairy solutions in \texorpdfstring{$ D= 4$}{D = 4}, \texorpdfstring{$ \mathcal{N} = 8$}{N = 8} gauged supergravity}
We now turn to analyze four dimensional hairy black holes, starting from a generalized consistent truncation of four--dimensional of $\mathcal{N}=8$ four dimensional supergravity \cite{Faulkner2010, Duff1999}:
$$
S  =  \frac{1}{8 \pi G_4} \int d^4x\, \sqrt{ - g} \left( \frac{R}{2} - \frac{1}{2}( \partial \varphi)^2 +  \cosh \left(\frac{ \sqrt{2} \varphi}{n} \right) + \cosh \left( \frac{ \sqrt{2} b \varphi}{n} \right) + \cosh  \left(\frac{ \sqrt{2} c\varphi}{n} \right) \right) \;, 
$$
where $b$ and $c$ are free parameters and:
$$
n  \equiv  \sqrt{1 + b^2 + c^2} \;.
$$
The terms in the potential are combined in such a way that $m^2 = - 2$. Accordingly, the scalar field decays at the AdS boundary as:
$$
\varphi = \frac{ \alpha}{r} + \frac{ \beta}{r^2} \;,
$$
and we consider boundary conditions of the form $ \beta = W _{, \alpha}$.

With usual boundary conditions $ \alpha = 0$, the dual field theory is defined on $ \mathbb{R} \times S^2$, and it has been conjectured to be the so--called ABJM theory \cite{Aharony2008b}. $ \alpha$ now represents the vev of a dimension one operator, and boundary conditions $ \beta = W_{, \alpha}$ induce a deformation of the boundary action:
$$
\alpha = \langle \mathcal{O} \rangle, \quad S \rightarrow S - \frac{R_{AdS}^2}{8 \pi G_4} \int dt\, d^2 \theta\, \sqrt{ \Gamma}\, W( \mathcal{O}) \;.
$$
Like in five--dimensions, one can obtain hairy black hole and hairy soliton solutions by numerically solving the static, spherically symmetric field equations. The ensemble of solutions of a given radius $ r_h$ determines a curve in the $ ( \alpha, \beta)$ plane, that we now parametrize as $ \beta _{r _{h}}( \alpha)$.  The scalar potential being even, we can restrict our analysis to solutions with $ \alpha >0$. As usual, when $ r_h \rightarrow 0$ one recovers the curve characterizing hairy solitons solutions, $ \beta_s( \alpha)$. Solving the scalar field equation on $AdS$ one can show that:
 $$
 \beta_s'( \alpha = 0) = - \frac{2}{ \pi} \;.
 $$
\begin{figure}[t!]
\begin{minipage}{\smallWidthLeft}
\includegraphics[width=\smallWidthRight]{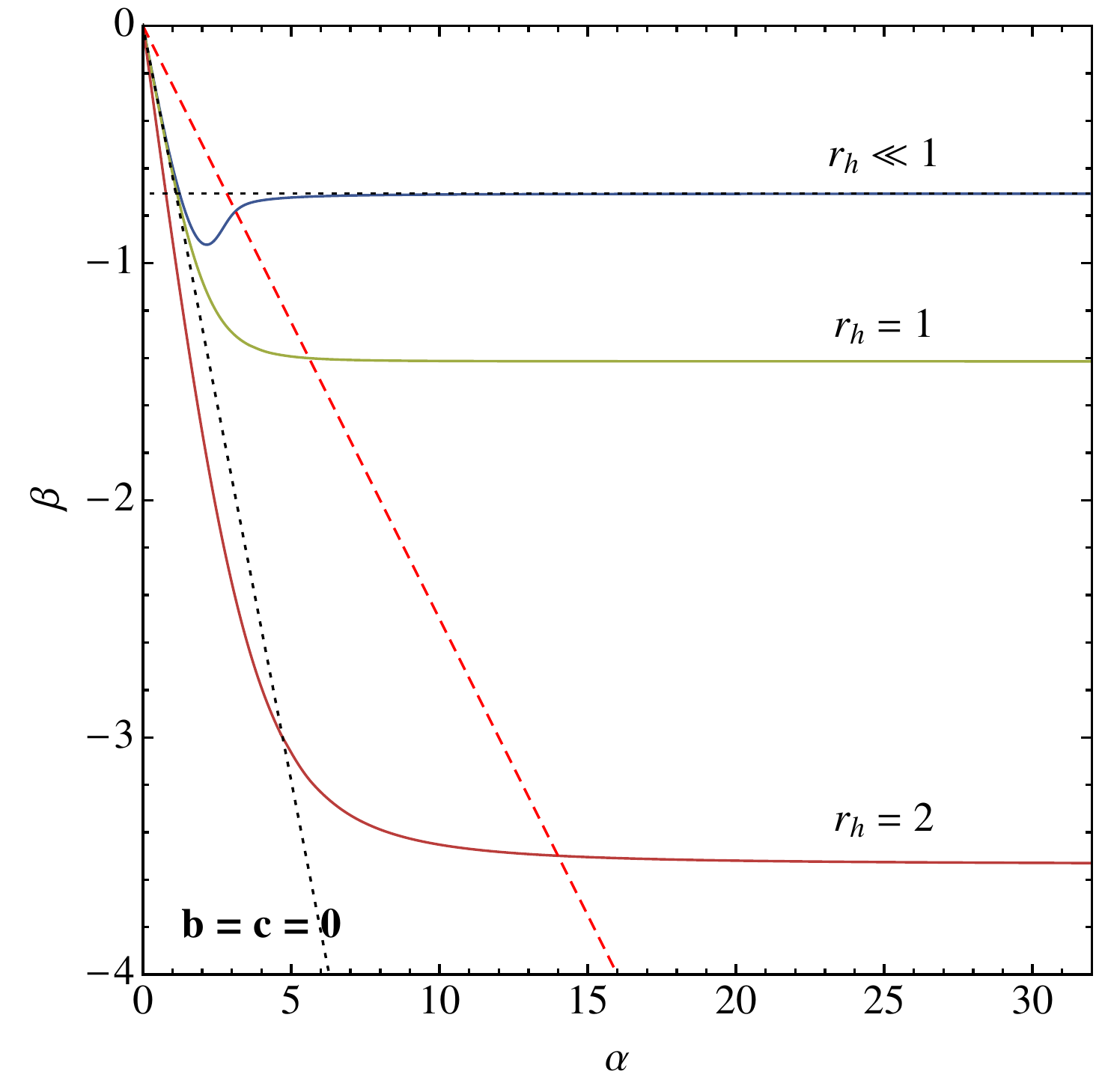}
\end{minipage}%
\begin{minipage}{\smallWidthRight}
\includegraphics[width=\smallWidthRight]{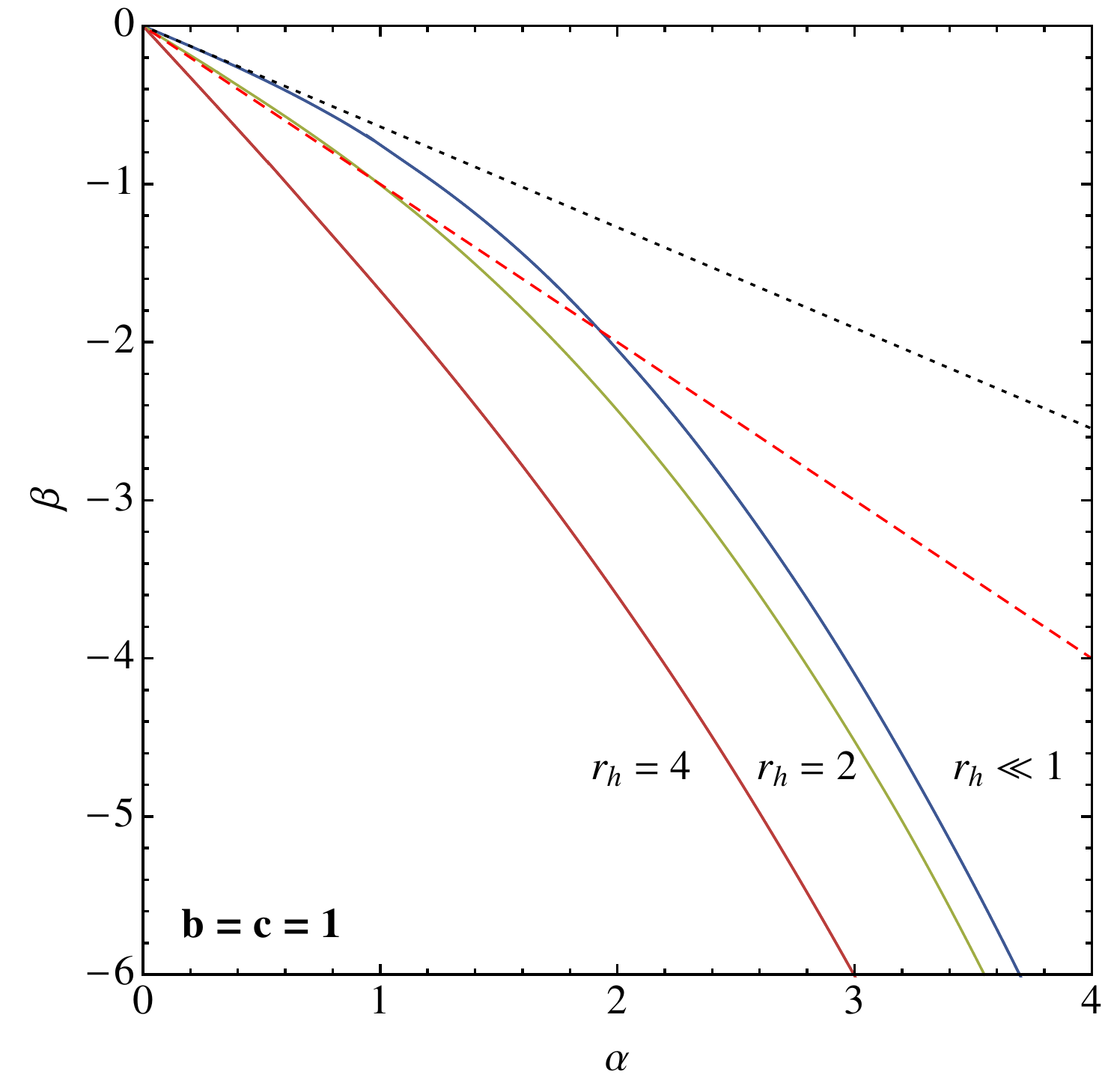}
\end{minipage}
\begin{minipage}{\smallWidthLeft}
\includegraphics[width=\smallWidthRight]{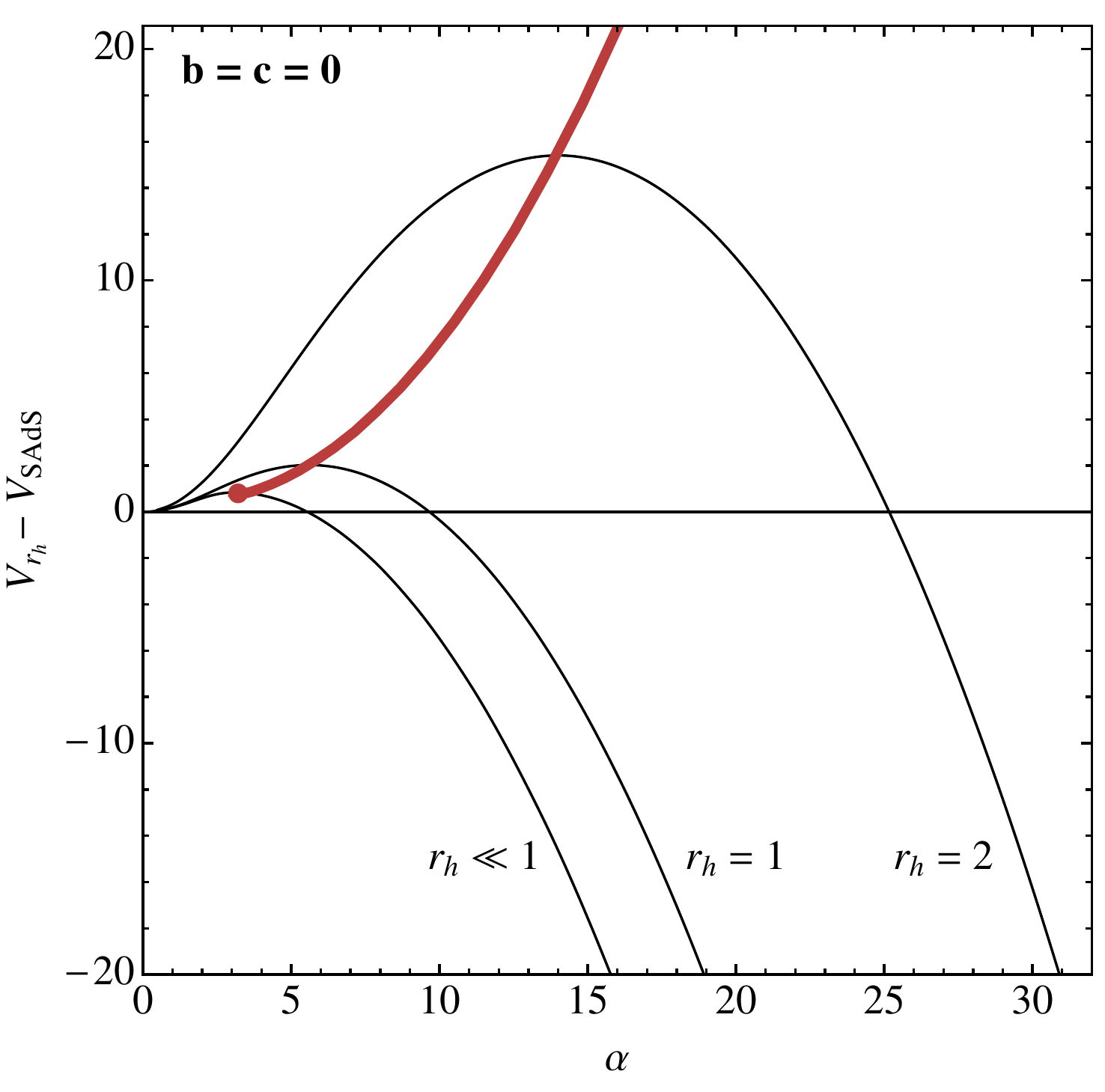}
\end{minipage}%
\begin{minipage}{\smallWidthRight}
\includegraphics[width=\smallWidthRight]{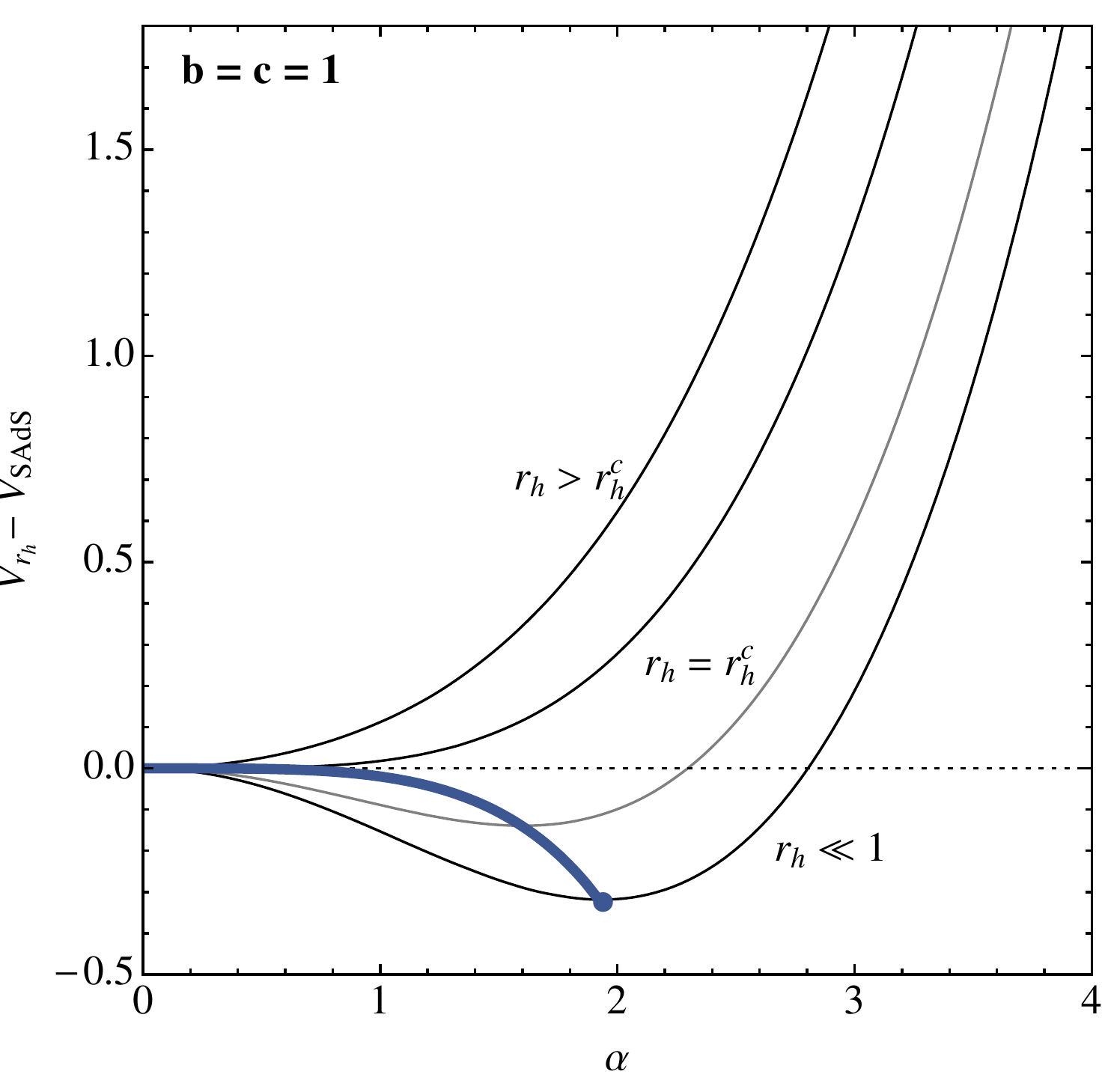}
\end{minipage}
\caption{ \small \label{fig:4DalphaBetaandfamilyb0c0andb1c1} In the top panels, some $ \beta _{r _{h}}( \alpha)$ curves for $b=c=0$ and $b=c=1$ respectively; the black, dotted line represents the slope of $ \beta_s$ at the origin, independent from $ b$ and $c$; the red, dashed lines represent $ \beta_f$ boundary conditions for $ f = 0.25$ and $f = 1$ respectively. In the bottom panels, we show the corresponding form of the constant radius effective potentials. For $b=c=0$, the hairy black holes corresponding to the maximum of $ \mathcal{V}_{r_h}$ become more and more hairy and massive, with respect to SAdS, as $r_h$ grows. For $ b=c=1$, the hairy black holes have lower mass than SAdS, until they disappear at $r_h=r_h^c$.}
\end{figure}%
At the non--linear level, the precise form of $ \beta_s$ clearly depends on the particular truncation. For generic values of $ b$ and $ c$ the soliton curve scales, at large $ \alpha$, as one would expect from dimensional analysis \cite{Faulkner2010}:
$$
\beta_s \simeq - s_c \, \alpha^2, \quad s_c>0 \;.
$$
According to the definition of $\mathcal{V}_s$ \eqref{eq:effPotsNeq}, this implies that the soliton effective potential  contains a positive definite term proportional to  $ \alpha^3$. This term insures a lower bound to the energy even in presence of boundary conditions that imply a perturbative instability of $AdS$ \cite{Faulkner2010,Ishibashi2003}:
\begin{eqnarray*}
W( \alpha) & = & - \frac{f}{2} \alpha^2, \quad f> \frac{ 2}{ \pi}\\
 \beta_f & = & - f \alpha \;.
\end{eqnarray*}
When $ f$ is large, the soliton effective potential takes the mexican hat form:
$$
8 \pi G_4 \, \mathcal{ V}_s \simeq - \frac{f}{2} \alpha^2 + \frac{s_c}{3} \alpha^3 \;.
$$
The hairy soliton sitting at the minimum of $\mathcal{ V}_s$ has negative mass. When the radius is increased, the slope of $ \beta_{r_h}$ at the origin becomes more and more negative and eventually $  \beta_{r_h}'( \alpha=0) < - f$ (see Figure \ref{fig:4DalphaBetaandfamilyb0c0andb1c1}, right column). Hence, one finds hairy black hole solutions compatible with $ \beta_f$ only for $r_h$ smaller than some critical value $r_h ^{c}$ or, equivalently, for temperatures smaller than some $T ^{c}$. Both $ r_h^{c}$ and $ T^{c}$ are increasing functions of $ f$, that can be evaluated analytically for large $f$:
\begin{equation}
r_h^c, T^c \propto f \;.
\end{equation}
In the picture provided by constant radius effective potentials, the non trivial minimum of $ \mathcal{V}_{r_h}$ disappears when $r_h$ approaches $r_h^{c}$ (see Figure \ref{fig:4DalphaBetaandfamilyb0c0andb1c1}). In the minimal generalization of the bulk action containing a $U(1)$ gauge field,  this is interpreted in the dual field theory as the continuous restoring,  at high temperature, of a spontaneously broken symmetry. The breaking is induced by the deformation corresponding to the boundary conditions, without the need to add a finite chemical potential  \cite{Faulkner2010b}.

The truncation obtained with $ b=c=0$ provides a completely different picture. As already noticed in \cite{Hertog2005b}, here the soliton curve approaches a constant value when $ \alpha$ goes to infinity, $ \beta \rightarrow \beta _{c} \simeq - 0.7071$. This asymptote corresponds to a mass term for $ \mathcal{O}$ in the soliton effective potential. Adopting $ \beta_f$ boundary conditions, the latter takes the asymptotic form:
\begin{equation} \label{eq:effPotb0c0}
8 \pi G_4 \, \mathcal{ V}_s \simeq | \beta _{c}| \alpha - \frac{f}{2} \alpha^2 \;.
\end{equation}
Therefore, one finds a hairy soliton compatible with $ \beta_f$ boundary conditions represented by the maximum of $ \mathcal{V}_s$ (see Figure \ref{fig:4DalphaBetaandfamilyb0c0andb1c1}, left column), that is now unbounded from below. At finite radius, the $ \beta_{r_h}$ curve converges to a constant value that, unlike in five dimensions, increases with $r_h$. Therefore, $ \beta_f$ boundary conditions admit hairy black holes of arbitrarily large radius.

For definiteness, we shall concentrate on the models defined by $ b=c=1$ and $b=c=0$, adopting $ \beta_f$ boundary conditions.

\subsection*{Truncation \texorpdfstring{$\mathbf{b = c = 0}$}{$b=c=0$}}
For $b=c=0$, the shape of the effective potential suggests again the presence of an instability of the hairy solutions under perturbations of the hair. Such instability has indeed been found \cite{Hertog2005c} with different but qualitatively similar boundary conditions, and appears a single pure imaginary frequency with positive imaginary part in the spectrum of normal perturbations. Like in five dimensions, one can again show that the frequency of such a mode must necessarily be pure imaginary. The proof follows the same line as before: near the boundary, the reduced perturbation variable behaves as follows:
 $$
 \Phi = A - B\, r ^{*} + O( r ^{* 2} ) \;,
 $$
 where now the boundary conditions impose $B = W_{, \alpha \alpha}( \alpha) A$.
 The last step of the proof \eqref{eq:proof} is replaced by:
$$
\textrm{Im}\, \left( \Phi ^{\star} \frac{ d \Phi}{d r ^{*}} \right)\, \stackrel{ r ^{*} \rightarrow 0 ^{-}}{\longrightarrow} \, \textrm{Im} \left( W_{, \alpha \alpha} |A|^2 \right) = 0 \;.
$$
The presence of this perturbative instability is confirmed by our numerical results: an pure imaginary frequency $ \omega = i \nu _{0}$, $ \nu_0 >0$ characterizes all the hairy solitons and black holes we have considered.
Again, a spurious frequency $ \tilde{ \omega} = i \tilde{ \nu}_0$, $ \tilde{ \nu_0} <0$ is also present, with $ \nu_0(r_h = 0) = \tilde{ \nu}_0(r_h = 0)$. 
\begin{figure}[t!]
\centering
\includegraphics[width=\hugeWidth]{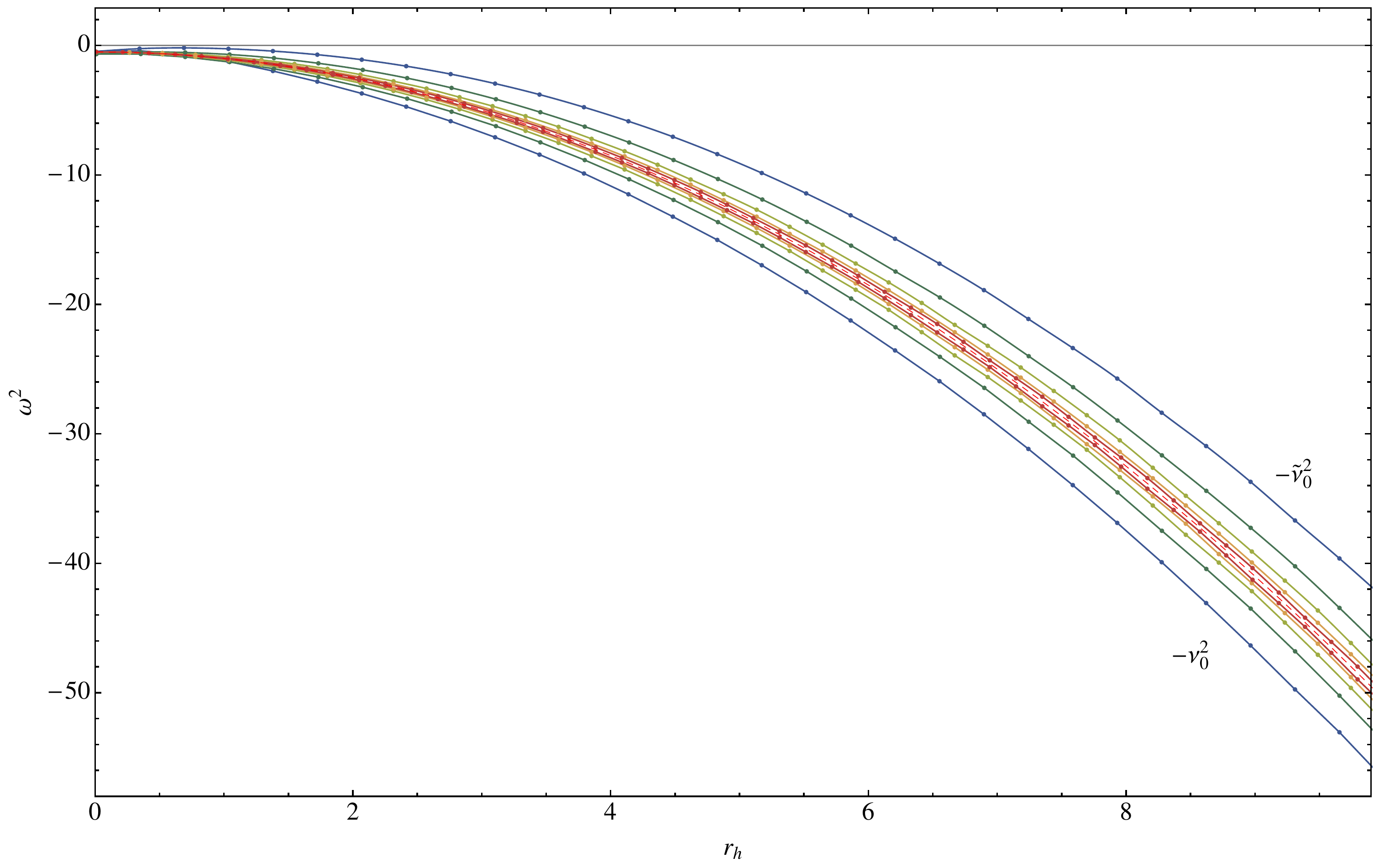}
\caption{ \small \label{fig:4Dmodesb0c0} \textbf{b=c=0}: Comparison between the squared frequency $ - \nu_0^2$ of the growing mode and the curvature of the constant entropy effective potential, $ \mathcal{V}_{r_h}$, for various values of $f$: from the outermost curves, $ f=0.5, \,0.25,\,0.125,\,0.0625,\,0.03125$. The red dashed line at the center represents the curvature of the effective potential $ \mathcal{V}_{r_h}$. In the large field limit $f \rightarrow 0$, one finds $ \nu_0^2 = \tilde{\nu}_0^2 = - \omega_{r_h}^2$.
}
\end{figure}%
When $ f$ goes to zero, both $ \nu_0^2(r_h)$ and $ \tilde{ \nu}_0^2(r_h)$ converge to a single universal function of the radius (see Figure \ref{fig:4Dmodesb0c0})\footnote{This feature also appears with the $AdS$--invariant boundary conditions $ \beta = - f \alpha^2$ considered in \cite{Hertog2005c}. This has probably not been recognized by the authors due to a background dependent gauge choice, $ \delta(r_h) = 0$: this implies that $ \delta$ has different values at the AdS boundary for different backgrounds, where times are measured in different units.}. This suggests that, like in five dimensions, in the limit of large condensates the frequency of these perturbations may be described by the curvature of the constant radius effective potential. Using the approximate expression \eqref{eq:effPotb0c0}, we see that the soliton solution is located at $ \alpha _{-} \propto f^{-1}$. Therefore, considering a boundary scalar $ \phi$ such that $ \alpha \propto \phi^2$, we find again:
$$
\frac{d^2 \mathcal{V}_s}{d \phi^2} \propto f^{0} \;.
$$
However, to make a quantitative comparison, we need to fix the normalization of $ \alpha$ in terms of $ \phi$.  Like in five dimension, we tentatively focus on the mass term appearing in the expression of the soliton potential at large $ \alpha$:
$$
8 \pi G_4 \, \mathcal{ V}_s \simeq | \beta _{c}| \alpha + W( \alpha) \;.
$$
As $ \alpha$ represents the vev of a dimension one operator $ \mathcal{O}$, and the dual field theory is defined on $ \mathbb{R} \times S^2$, we try to identify such term with the conformal coupling of $ \mathcal{O}$ to the curvature of $ S^2$ \cite{Hertog2005b}. Neglecting the non--abelian structure of the field theory, we again consider a toy model free theory for a single scalar field:
\begin{eqnarray*}
\mathcal{O} & = & c\, \phi^2 \;,\\
\mathcal{ V}_s & = & \frac{| \beta _{c}| \alpha + W( \alpha)}{8 \pi G_4} = \frac{1}{8} \phi^2 + \frac{ W( \alpha)}{8 \pi G_4} \;,\\
c & = & \frac{\pi G_4}{| \beta_c|} \;,
\end{eqnarray*}
where $ \phi$ is supposed to be canonically normalized in the dual field theory. With this prescription, we can compare the quasi--normal frequency $ -\nu_0^2$, as a function of $r_h$, with the curvature of the constant radius effective potential $ \mathcal{V}_{r_h}$ at its extrema, computed in $ \phi$ variables:
$$
\omega _{r_h}^2  \equiv  \frac{d^2 \mathcal{V}_{r_h}}{d \phi^2} = 4\, c \,\alpha \frac{d^2 \mathcal{V}_{r_h}}{d \alpha^2} = \frac{ \alpha}{2 | \beta_c|} \left( - \frac{d \beta_{r_h}}{d \alpha} + W_{, \alpha \alpha} \right) \;.
$$
Once again, $ \omega_{r_h}^2$ can be easily evaluated numerically, once the hairy black holes are found. In Figure \ref{fig:4Dmodesb0c0}, we show the numerical results for the agreement between $ \omega_{r_h}^2$ and $ -\nu_0^2$, $- \tilde{ \nu}_0^2$. While $f$ decreases, the agreement gets better and better, and $ \nu_0 \simeq \tilde{ \nu}_0$ as one would expect in the absence of dissipation. Therefore, like in five dimensions, this truncation exhibits a large field regime in which the effective soliton potential can be consistently identified with the multi trace deformation plus a mass term that couples the dual operator with the curvature of $S^2$.
\begin{figure}[t!]
\includegraphics[width=\hugeWidth]{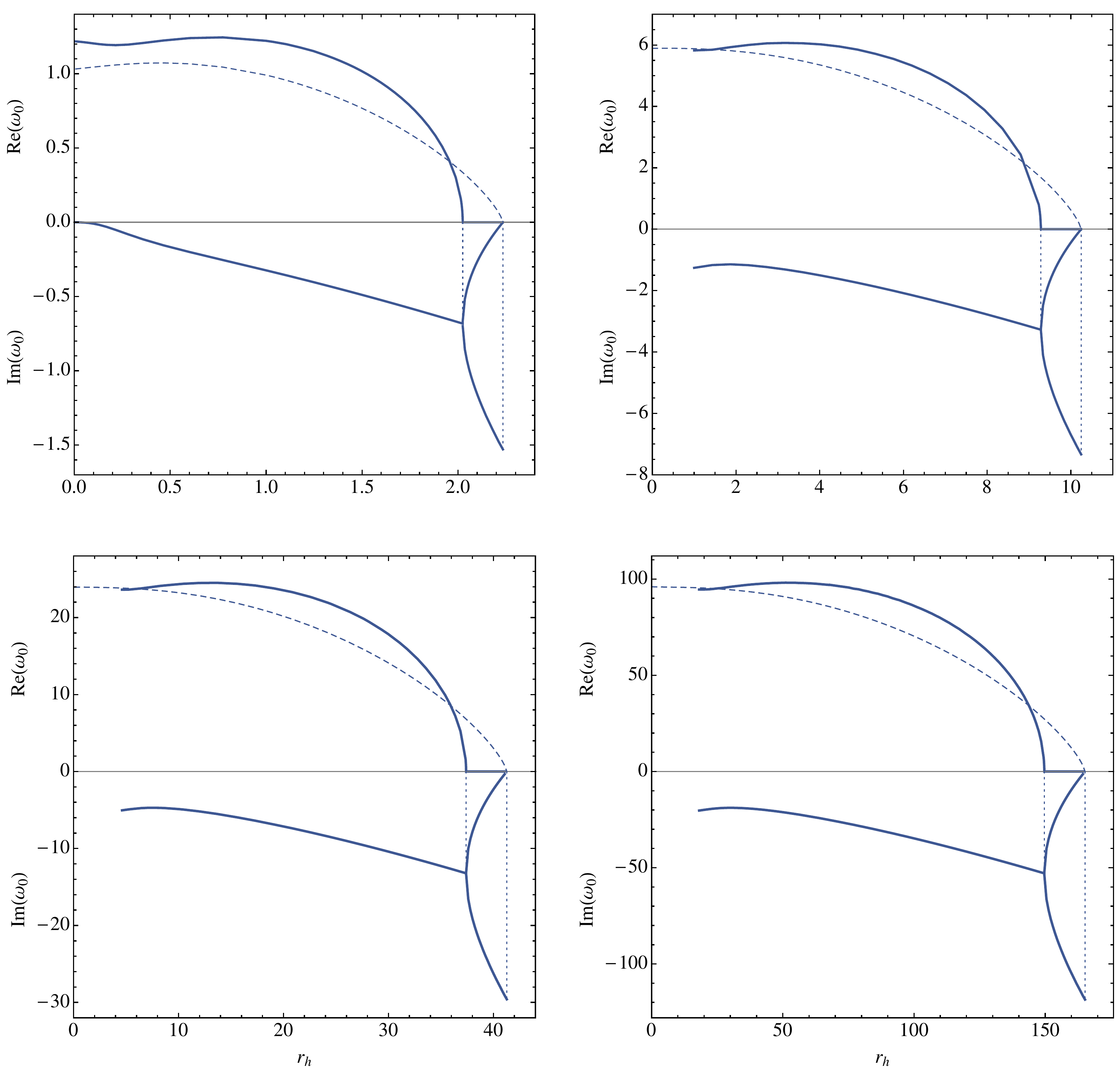}
\caption{ \small \label{fig:4Dconjectureb1c1} \textbf{b = c = 1}: real and imaginary part of the least damped quasi--normal frequency of hairy black holes, for boundary conditions with $f=1$ (top left panel), $f=4$ (top right), $f=16$ (bottom left), $f = 64$ (bottom right). The imaginary part is also shown. Solid lines represent data from quasi--normal frequencies, while dashed lines represent the square root of the second derivative of the radius--dependent constant entropy effective potential, evaluated at its two extrema. The frequencies at small radius, relatively to $ f$, turned out to be above the confidence limit of our numerical computation (see discussion in Section 2), and are not show. In the large field limit $f \rightarrow \infty$ the real and imaginary part of $ \omega_0$, as functions of $f^{-1}r_h$, globally scale as $ f$. Therefore, $ \omega_0$ is still associated with a large dissipation.}
\end{figure}%
We have also considered ``regularized'' boundary conditions that recall the one adopted in five dimensions \cite{Hertog2005a}:
$$
\beta_{f, \eta} = - f \alpha + f^2 \eta \alpha^2, \quad \eta >0 \;.
$$
The picture we have obtained is essentially the same: the regularizing term produces a new branch of perturbatively stable black holes that have negative mass at small radius. They are characterized by a quasi--normal frequency $ \omega_0$ that, in the small $f$ limit, becomes stable ($ \textrm{Im}( \omega_0) \rightarrow 0$) and whose real part is well predicted by the curvature of $ \mathcal{V}_{r_h}$.

\subsection*{Truncation \texorpdfstring{$\mathbf{b = c = 1}$}{$b=c=1$}}
For $b=c=1$, the shape of the effective potential with $ \beta _f$ boundary conditions strongly recalls the one we encountered in the study five dimensional hairy black holes. There, the critical radius $ r_h^c$ corresponded to the joining point of the two extrema of $ \mathcal{V}_{r_h}$, both located at $ \beta > 0$ and corresponding to unstable/stable hairy black holes. Here, SAdS black holes play the role of the unstable hairy black holes. Therefore, we expect to find a special frequency $ \omega_0$, in the spectrum of quasi--normal perturbations, that vanishes for $r _h \rightarrow r_h^c$. 
However, the situation is rather different from the previous cases. The large $f$ form of $ \mathcal{V}_s$:
$$
8 \pi G_4 \, \mathcal{ V}_s \simeq - \frac{f}{2} \alpha^2 + \frac{s_c}{3} \alpha^3 \;,
$$
is dominated, when $ \alpha$ is large, by the term $ s_c \alpha^3$, that is not present in the deformation $W( \alpha)$ and cannot be identified as a free term of the undeformed field theory action. In fact, the behavior of the quasi--normal frequency $ \omega_0$ shows a qualitative resemblance with the five dimensional case (for comparison, Figures \ref{fig:5Dconjecture} and \ref{fig:4Dconjectureb1c1}). However, the limit of large condensates is very different: when $ f$ grows, one recovers the scaling suggested by dimensional analysis:
\begin{equation} \label{eq:scalingb1c1}
\omega_0  \propto  f^{-1} \;,
\end{equation}
both for the real and imaginary part of $ \omega_0$. In particular, $ \omega_0$ reaches the imaginary axis at a value of $r_h$ that scales as $f^{-1}$ and is significantly smaller than $ r_h^c$. Therefore, even if the scaling \eqref{eq:scalingb1c1} naturally characterizes the curvature of $ \mathcal{V}_s$ for the same dimensional reason, we cannot expect to find a quantitative agreement between $ \textrm{Re}( \omega_0^2)$ and $ \omega_{r_h}^2$.

\section*{Conclusions}
We have given detailed numerical results for the $l=0$ scalar quasi--normal frequencies of four and five dimensional AdS black holes with scalar hair and generalized boundary conditions. The computation has been performed by taking in account the full backreaction of the scalar field on the metric.

We have identified models in which, at large values of the ``hair'', the difference between the effective potential $ \mathcal{V}_s$ and the deformation of the dual theory can be naturally identified as a conformal mass term in the field theory action. As a consequence, the normalization of the operator with respect to a canonically normalized scalar can be numerically obtained from the soliton data. The corresponding hairy black holes possess a quasi--normal  frequency $ \omega_0$ which is quasi stable ($| \textrm{Im}( \omega_0)/ \omega_0| \ll 1$) and coincides with the curvature of the effective potential at fixed entropy $ \mathcal{V}_{r_h}$, provided the condensate is large enough. In the dual field theory, the corresponding mode is associated to homogenous perturbations of the condensate that undergo very weak dissipation. A heuristic interpretation for the presence such mode is that higher derivatives of $ \mathcal{V}_{r_h}$, corresponding to self--couplings of the perturbations, vanish in the limit of large condensates. Oppositely, the frequency of higher overtones mantain a finite imaginary part.

Boundary conditions have been considered for which two branches of stable and unstable hairy black holes join at a critical point, where the curvature of $ \mathcal{V}_{r_h}$ vanishes. At the critical point, the mode mentioned above has zero frequency and describes the transition between the two branches. For smaller values of the radius, its frequency turns to pure imaginary values before bifurcating into two frequencies with non zero real part.

Finally, we have discussed a four dimensional model in which $ \mathcal{V}_s$ contains an interaction term deduced from the supergravity data, which is not present in the deformation. Here, the imaginary part of the least damped frequency does not vanish in the limit of large condensates, and the constant entropy effective potential describes its real part only at the level of the scaling with the parameters of the deformation.  This quantitatively different behavior may be interpreted by the fact that, unlike in the first class of models, some higher derivatives of $ \mathcal{V}_s$ do not vanish in the limit of large condensates.

\section*{Acknowledgements}
We would like to thank Thomas Hertog for many helpful discussions during all the stages of this project.

\bibliographystyle{utphys}
\bibliography{qnModesDesignerGravity}

\providecommand{\href}[2]{#2}\begingroup\raggedright\begin{thebibliography}{10}

\bibitem{Horowitz2000}
G.~T. {H}orowitz and V.~E. {H}ubeny, ``{Quasinormal modes of AdS black holes
  and the approach to thermal equilibrium},''
  \href{http://dx.doi.org/10.1103/PhysRevD.62.024027}{{\em {P}hys. {R}ev.} {\bf
  D62} (2000)  024027},
\href{http://arxiv.org/abs/hep-th/9909056}{{\tt arXiv:hep-th/9909056}}.
%%CITATION = HEP-TH/9909056;%%.

\bibitem{Birmingham2002}
D.~{B}irmingham, I.~{S}achs, and S.~N. {S}olodukhin, ``{Conformal field theory
  interpretation of black hole quasi- normal modes},''
  \href{http://dx.doi.org/10.1103/PhysRevLett.88.151301}{{\em {P}hys. {R}ev.
  {L}ett.} {\bf 88} (2002)  151301},
\href{http://arxiv.org/abs/hep-th/0112055}{{\tt arXiv:hep-th/0112055}}.
%%CITATION = HEP-TH/0112055;%%.

\bibitem{Son2002}
D.~T. {S}on and A.~O. {S}tarinets, ``{Minkowski-space correlators in AdS/CFT
  correspondence: Recipe and applications},'' {\em {JHEP}} {\bf 09} (2002)
  042,
\href{http://arxiv.org/abs/hep-th/0205051}{{\tt arXiv:hep-th/0205051}}.
%%CITATION = HEP-TH/0205051;%%.

\bibitem{Hertog2004d}
T.~{H}ertog and G.~T. {H}orowitz, ``{Towards a big crunch dual},''
  \href{http://dx.doi.org/10.1088/1126-6708/2004/07/073}{{\em {JHEP}} {\bf 07}
  (2004)  073},
\href{http://arxiv.org/abs/hep-th/0406134}{{\tt arXiv:hep-th/0406134}}.
%%CITATION = HEP-TH/0406134;%%.

\bibitem{Craps2007}
B.~{C}raps, T.~{H}ertog, and N.~{T}urok, ``{Quantum Resolution of Cosmological
  Singularities using AdS/CFT},''
\href{http://arxiv.org/abs/0712.4180}{{\tt arXiv:0712.4180 [hep-th]}}.
%%CITATION = 0712.4180;%%.

\bibitem{Faulkner2010b}
T.~{F}aulkner, G.~T. {H}orowitz, and M.~M. {R}oberts, ``{Holographic quantum
  criticality from multi-trace deformations},''
\href{http://arxiv.org/abs/1008.1581}{{\tt arXiv:1008.1581 [hep-th]}}.
%%CITATION = 1008.1581;%%.

\bibitem{Hertog2005b}
T.~{H}ertog and G.~T. {H}orowitz, ``{Designer gravity and field theory
  effective potentials},''
  \href{http://dx.doi.org/10.1103/PhysRevLett.94.221301}{{\em {P}hys. {R}ev.
  {L}ett.} {\bf 94} (2005)  221301},
\href{http://arxiv.org/abs/hep-th/0412169}{{\tt arXiv:hep-th/0412169}}.
%%CITATION = HEP-TH/0412169;%%.

\bibitem{Hertog2005c}
T.~{H}ertog and K.~{M}aeda, ``{Stability and thermodynamics of AdS black holes
  with scalar hair},'' \href{http://dx.doi.org/10.1103/PhysRevD.71.024001}{{\em
  {P}hys. {R}ev.} {\bf D71} (2005)  024001},
\href{http://arxiv.org/abs/hep-th/0409314}{{\tt arXiv:hep-th/0409314}}.
%%CITATION = HEP-TH/0409314;%%.

\bibitem{Battarra2010}
L.~{B}attarra and T.~{H}ertog, ``{Particle Production near an AdS Crunch},''
  \href{http://dx.doi.org/10.1007/JHEP12(2010)017}{{\em {JHEP}} {\bf 12} (2010)
   017},
\href{http://arxiv.org/abs/1009.0992}{{\tt arXiv:1009.0992 [hep-th]}}.
%%CITATION = 1009.0992;%%.

\bibitem{Witten2001}
E.~{W}itten, ``{Multi-trace operators, boundary conditions, and AdS/CFT
  correspondence},''
\href{http://arxiv.org/abs/hep-th/0112258}{{\tt arXiv:hep-th/0112258}}.
%%CITATION = HEP-TH/0112258;%%.

\bibitem{Vecchi2010b}
L.~{V}ecchi, ``{Multitrace deformations, Gamow states, and Stability of
  AdS/CFT},''
\href{http://arxiv.org/abs/1005.4921}{{\tt arXiv:1005.4921 [hep-th]}}.
%%CITATION = 1005.4921;%%.

\bibitem{Balasubramanian1999}
V.~{B}alasubramanian and P.~{K}raus, ``{A stress tensor for anti-de Sitter
  gravity},'' \href{http://dx.doi.org/10.1007/s002200050764}{{\em {C}ommun.
  {M}ath. {P}hys.} {\bf 208} (1999)  413--428},
\href{http://arxiv.org/abs/hep-th/9902121}{{\tt arXiv:hep-th/9902121}}.
%%CITATION = HEP-TH/9902121;%%.

\bibitem{Hertog2005a}
T.~{H}ertog and G.~T. {H}orowitz, ``{Holographic description of AdS
  cosmologies},'' \href{http://dx.doi.org/10.1088/1126-6708/2005/04/005}{{\em
  {JHEP}} {\bf 04} (2005)  005},
\href{http://arxiv.org/abs/hep-th/0503071}{{\tt arXiv:hep-th/0503071}}.
%%CITATION = HEP-TH/0503071;%%.

\bibitem{Faulkner2010}
T.~{F}aulkner, G.~T. {H}orowitz, and M.~M. {R}oberts, ``{New stability results
  for Einstein scalar gravity},''
  \href{http://dx.doi.org/10.1088/0264-9381/27/20/205007}{{\em {C}lass.
  {Q}uant. {G}rav.} {\bf 27} (2010)  205007},
\href{http://arxiv.org/abs/1006.2387}{{\tt arXiv:1006.2387 [hep-th]}}.
%%CITATION = 1006.2387;%%.

\bibitem{Zhidenko2006}
A.~{Z}hidenko, ``{Quasi-normal modes of the scalar hairy black hole},''
  \href{http://dx.doi.org/10.1088/0264-9381/23/9/024}{{\em {C}lass. {Q}uant.
  {G}rav.} {\bf 23} (2006)  3155--3164},
\href{http://arxiv.org/abs/gr-qc/0510039}{{\tt arXiv:gr-qc/0510039}}.
%%CITATION = GR-QC/0510039;%%.

\bibitem{Kaminski2010a}
M.~{K}aminski {\em et al.}, ``{Quasinormal modes of massive charged flavor
  branes},'' \href{http://dx.doi.org/10.1007/JHEP03(2010)117}{{\em {JHEP}} {\bf
  03} (2010)  117},
\href{http://arxiv.org/abs/0911.3544}{{\tt arXiv:0911.3544 [hep-th]}}.
%%CITATION = 0911.3544;%%.

\bibitem{Gunaydin1985d}
M.~{G}unaydin, L.~J. {R}omans, and N.~P. {W}arner, ``{Gauged N=8 Supergravity
  in Five-Dimensions},''
\href{http://dx.doi.org/10.1016/0370-2693(85)90361-2}{{\em {P}hys. {L}ett.}
  {\bf B154} (1985)  268}.
%%CITATION = PHLTA,B154,268;%%.

\bibitem{Duff1999}
M.~J. {D}uff and J.~T. {L}iu, ``{Anti-de Sitter black holes in gauged N = 8
  supergravity},'' \href{http://dx.doi.org/10.1016/S0550-3213(99)00299-0}{{\em
  {N}ucl. {P}hys.} {\bf B554} (1999)  237--253},
\href{http://arxiv.org/abs/hep-th/9901149}{{\tt arXiv:hep-th/9901149}}.
%%CITATION = HEP-TH/9901149;%%.

\bibitem{Aharony2008b}
O.~{A}harony, O.~{B}ergman, D.~L. {J}afferis, and J.~{M}aldacena, ``{N=6
  superconformal Chern-Simons-matter theories, M2-branes and their gravity
  duals},'' \href{http://dx.doi.org/10.1088/1126-6708/2008/10/091}{{\em {JHEP}}
  {\bf 10} (2008)  091},
\href{http://arxiv.org/abs/0806.1218}{{\tt arXiv:0806.1218 [hep-th]}}.
%%CITATION = 0806.1218;%%.

\bibitem{Ishibashi2003}
A.~{I}shibashi and R.~M. {W}ald, ``{Dynamics in non globally hyperbolic static
  spacetimes. II: General analysis of prescriptions for dynamics},''
  \href{http://dx.doi.org/10.1088/0264-9381/20/16/318}{{\em {C}lass. {Q}uant.
  {G}rav.} {\bf 20} (2003)  3815--3826},
\href{http://arxiv.org/abs/gr-qc/0305012}{{\tt arXiv:gr-qc/0305012}}.
%%CITATION = GR-QC/0305012;%%.

\end{thebibliography}\endgroup

\end{document}